\documentclass{aa}
\input psfig.tex
\begin{document}

   \thesaurus{06     % A&A Section 6: Form. struct. and evolut. of stars
              (11.03.1;  % Galaxies: clusters,
               11.09.2;  % Galaxies: interaction,
               11.05.2;  % Galaxies: evolution,
               08.06.2)} % Stars: formation.
   \title{An H$_\alpha$ Catalogue of Galaxies in Hickson Compact Groups. I. The Sample}

%   \subtitle{I. Overviewing the $\kappa$-mechanism}

   \author{P. Severgnini, 
          \inst{1,4}
%          \and
       	  B. Garilli \inst{1,}
	  P. Saracco \inst{2}
	   \and
	  G. Chincarini\inst{2,3}
          }

   \offprints{Paola Severgnini}

   \institute{$^1$ IFCTR/CNR, Via Bassini, 15, 20133 Milano, Italy \\
%              email: 
%         \and
              $^2$ Osservatorio Astronomico di Brera-Merate, Milano, Italy \\
	      $^3$  Dip. di Fisica, Univ. Milano, Via Celoria, 16, 20133 Milano, Italy \\
	      $^4$ Univ. degli Studi di Firenze, Dip. Astronomia e Scienza dello Spazio, Via Fermi 5, 50125 Firenze (Arcetri), Italy.\\
	      email: paola@ifctr.mi.cnr.it; bianca@ifctr.mi.cnr.it; saracco@merate.mi.astro.it; guido@merate.mi.astro.it}
%             \thanks{The university of heaven temporarily does not
%                     accept e-mails}

   \date{Received ......; accepted ..........}

 \titlerunning{$H_\alpha$ Catalogue of HCG Galaxies}  
 \authorrunning{P. Severgnini et al.}
 \maketitle

\maketitle

   \begin{abstract}
We present H$_\alpha$ photometry for a sample of 95 galaxies in 
Hickson Compact Groups  obtained from  observations of 31 groups.
The Catalogue lists isophotal and adaptive aperture (Kron aperture) flux measurements for about 75$\%$ of the accordant galaxies inside the observed HCGs, 22 out of which are upper limits.
Non standard data reduction procedures have been used to obtain the 
continuum subtracted H$_\alpha$ images for each HCG of the target sample. 
Flux calibration has also been performed in order to obtain  H$_\alpha$ luminosities for the whole sample.
Both the data reduction and calibration procedures are carefully 
described in this paper.
The new data listed in this Catalogue are of great importance in understanding the star formation rate inside HCG galaxies and in giving new insights on its dependence on galaxy interactions.

   \end{abstract}

%
%________________________________________________________________
\section{Introduction}
Hickson Compact Groups (hereafter HCGs; Hickson \cite{Hic}; 
Hickson \cite{Hik}) are small systems of several galaxies (four or more) in an apparent close proximity in the sky.
The debate on their physical reality as bounded systems is still open. A possibility exists that only a part of the  sample of HCGs are bound systems and/or that HCG  dynamical evolution depends on their environments. Important informations about their real nature could be obtained by studying the rate of merger and interaction between their galaxies. The studies carried out so far  agree with the view of a low merging rate inside HCGs with respect to undisrupted systems of galaxies (Zepf et al. \cite{Zep:Whi}, Zepf \cite{Zep}). On the other hand it is not so clear which is the fraction of interacting galaxies in HCGs: photometric and spectroscopic studies  (Rubin et al. \cite{Rub}; Mendes de Oliveira et al. \cite{Men:Ama}; Moles et al. \cite{Mol:del};  Mendes de Oliveira et al. \cite{Men:Hic}; Vilchez \& Iglesias Paramo \cite{Vil:Para}; Vilchez \& Iglesias Paramo \cite{Vil:Parb}; Iglesias Paramo \& Vilchez \cite{Par:Vil}) have often given contradictory results. 
It is expected that interaction and merger phenomena strongly affect the star formation rate ({\em SFR}) of galaxies. 
In particular, interacting  galaxies should show an higher star formation rate than  field galaxies.
Thus the study of star formation of galaxies in HCGs gives important clues about the interaction and merger phenomena inside them.
Powerful tools to investigate on the star formation activity are the ionization lines emitted by the heated gas surrounding the regions of star formation.
The H$_\alpha$ emission line at 6563 {\AA} can be used as a quantitative and spatial tracer of the rate of massive ($\geq$ 10 M$_\odot$) and therefore recent ($\leq 10^7$ years) star formation (Kennicutt \cite{ken}; Ryder \& Dopita \cite{Ryd:Dop}), unlike the color indexes in the $U,B,V$ filters, that give indications about the past star formation  ($> 10^8$ years).
Therefore, by  knowing the  H$_\alpha$ emission of the HCG galaxies it is  possible in principle to carry out important investigations about the  present merger and interaction events in these systems. 
Up to now, only Rubin et al.  (\cite{Rub}) and more recently  Vilchez \& Iglesias Paramo (\cite{Vil:Para}) have collected significant samples of  H$_\alpha$ data on HCG galaxies.
They published H$_\alpha$ emission-line images respectively for 14 and 16 HCGs. While Vilchez \& Iglesias Paramo (\cite{Vil:Para}) estimate the H$_\alpha$ flux for each of the 63 galaxies of their sample, Rubin et al. do not use flux calibrated and they take into account a sample constituted by disk galaxies only.
H$_\alpha$ data for the galaxies of single groups have been also obtained by  Valluri \& Anupama (\cite{Val:Anu}), Mendes de Oliveira et al. (\cite{Men:Pla}) and Plana et al. (\cite{Pla:Men}).
Valluri \& Anupama presented H$_\alpha$ calibrated data  for the galaxies  of HCG62 and Mendes de Oliveira et al. and Plana et al. reported kinematic observations of H$_\alpha$ emission respectively for four late-type galaxies of HCG16 and for two early-type galaxies and one disk system of HCG90.

With the aim to obtain quantitative informations about the H$_\alpha$ emission of HCGs galaxies we have observed 31 HCG in narrow-band interferometric filters deriving H$_\alpha$ calibrated fluxes for 95 galaxies, 22 out of which are upper limits. In this paper we present  the catalogue containing these H$_\alpha$ data.\\
We first describe the sample and the observations in $\S$ 2 and in $\S$ 3. In $\S$ 4 and $\S$ 5 we present the data reduction and calibration procedures used. The Zero Point correction, Galactic and Internal extinction corrections applied to the fluxes are described in $\S$ 6, while  $\S$ 7 contains the photometric error derivation. 
In $\S$ 8 we present the H$_\alpha$ Catalogue of Galaxies, while in $\S$ 9 we derive the star formation rate for the whole sample. Finally we briefly discuss some of the observed groups in $\S$ 10.

%%%%%%%%%%%%%%%%%%%%%%%%%%%%%%% THE SAMPLE %%%%%%%%%%%%%%%%
\section{The Sample}
The 100 compact groups catalogue  by Hickson (\cite{Hic}) has been revised  by Hickson, Kindl \& Auman  (\cite{Hic:Kin}) and then by Hickson et al. (\cite{Hic:Men}). By adding a radial velocity criterion Hickson et al. (\cite{Hic:Men}) were able to reject probable non member galaxies. The resulting sample consists of 92 groups each containing three or more "accordant" members, which have radial velocities differing by no more than 1000 km~s$^{-1}$ from the median velocity of the group. Our sample has been drawn from this latest catalog.\\
In this paper, the result of the data reduction and calibration of H$_\alpha$ CCD images are presented for 31 Hickson Compact Groups. The remaining 61 HCGs were not in our sample because the adequate H$_\alpha$ interferometric filters were not available during the observations. This is the only criterion used to select the observed groups.\\
The redshift of observed groups is in the range 0.005$ \leq z \leq$ 0.07 (P. Hickson et al., \cite{Hic:Men}) and their distribution is shown in Figure 1 (the width of each bin is 0.01).
\begin{figure}[h]
\centerline{\psfig{figure=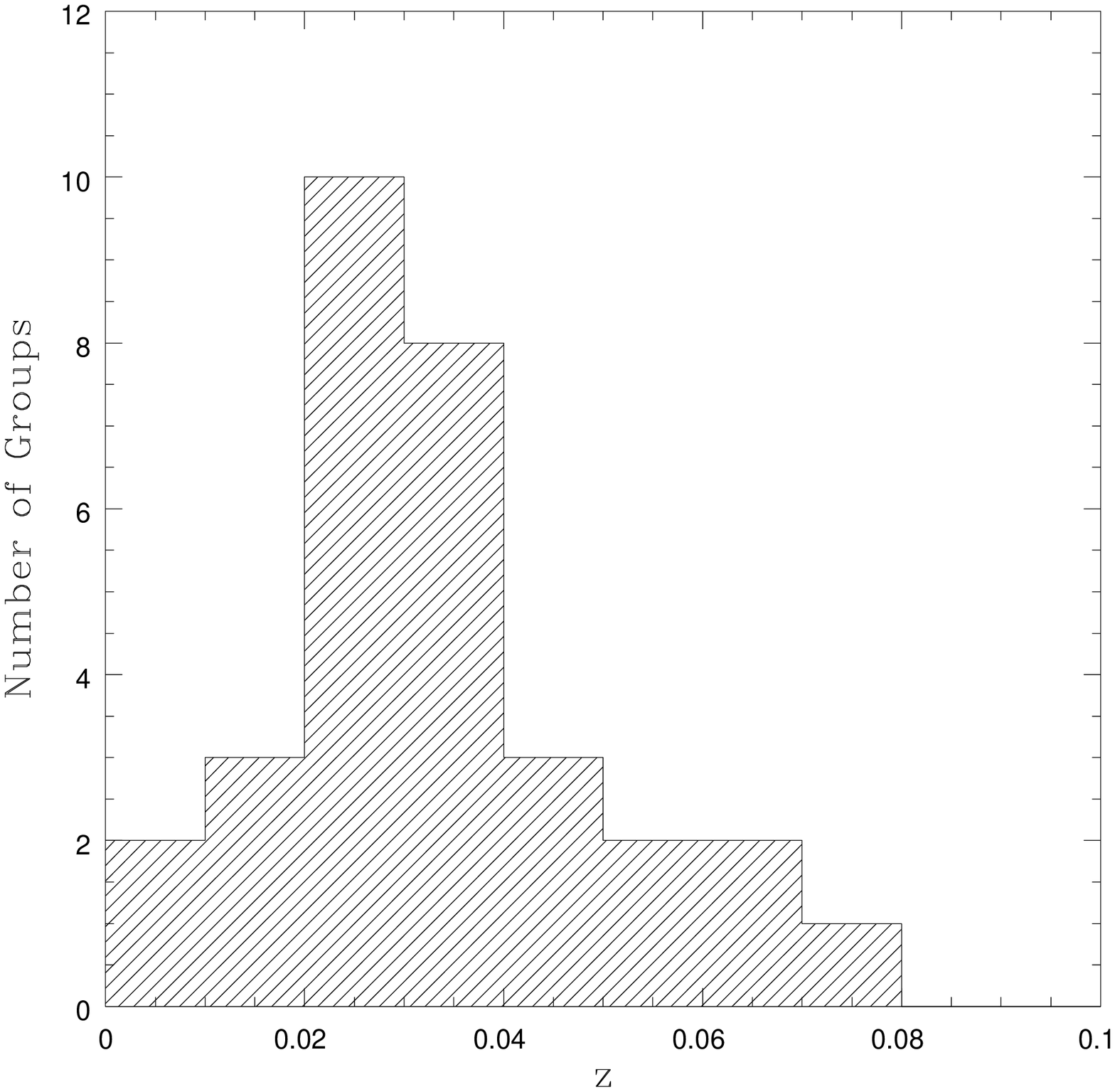,height=90mm,bbllx=70mm,bblly=60mm,bburx=140mm,bbury=250mm}}
\bigskip
\centerline{{\bf Figure 1:} Redshift distribution of the 31 observed groups.}
\end{figure}

\vskip 0.2truecm
Table 1 lists the observed HCGs as follows:\\
Col.1: Name of the groups according to Hickson's catalogue;\\
Col.2: 1950 right ascension (R.A.) of the centroid of the member galaxies;\\
Col.3: 1950 declination (Dec.) of the centroid of the member galaxies;\\
Col.4: Number of accordant members of the group;\\
Col.5: Velocity dispersion of the group: $\sigma_v$ $(Km~s^{-1})$;\\
Col.6: Mean Redshift of the group.\\

The sample is composed by 134 galaxies, 127 out of which have been observed. 52$\%$ of them are Ellipticals and Lenticulars and the remaining 48$\%$ are Spirals and Irregulars. For each observed galaxy we report in Table 2  the heliocentric radial velocity {\em V} in units of $Km~s^{-1}$, the total magnitude in the photografic band $B_T$, corrected for internal and galactic   extinction and the Hubble type, as in Hickson \cite{Hik}. Galaxies are named with the number of HCG plus letter of galaxy itself.\\
The distribution of the total $B_T$ magnitude of the observed galaxies is shown in Figure 2 (the width of each bin is 0.5 magnitude).

\begin{figure}[h]
\centerline{\psfig{figure=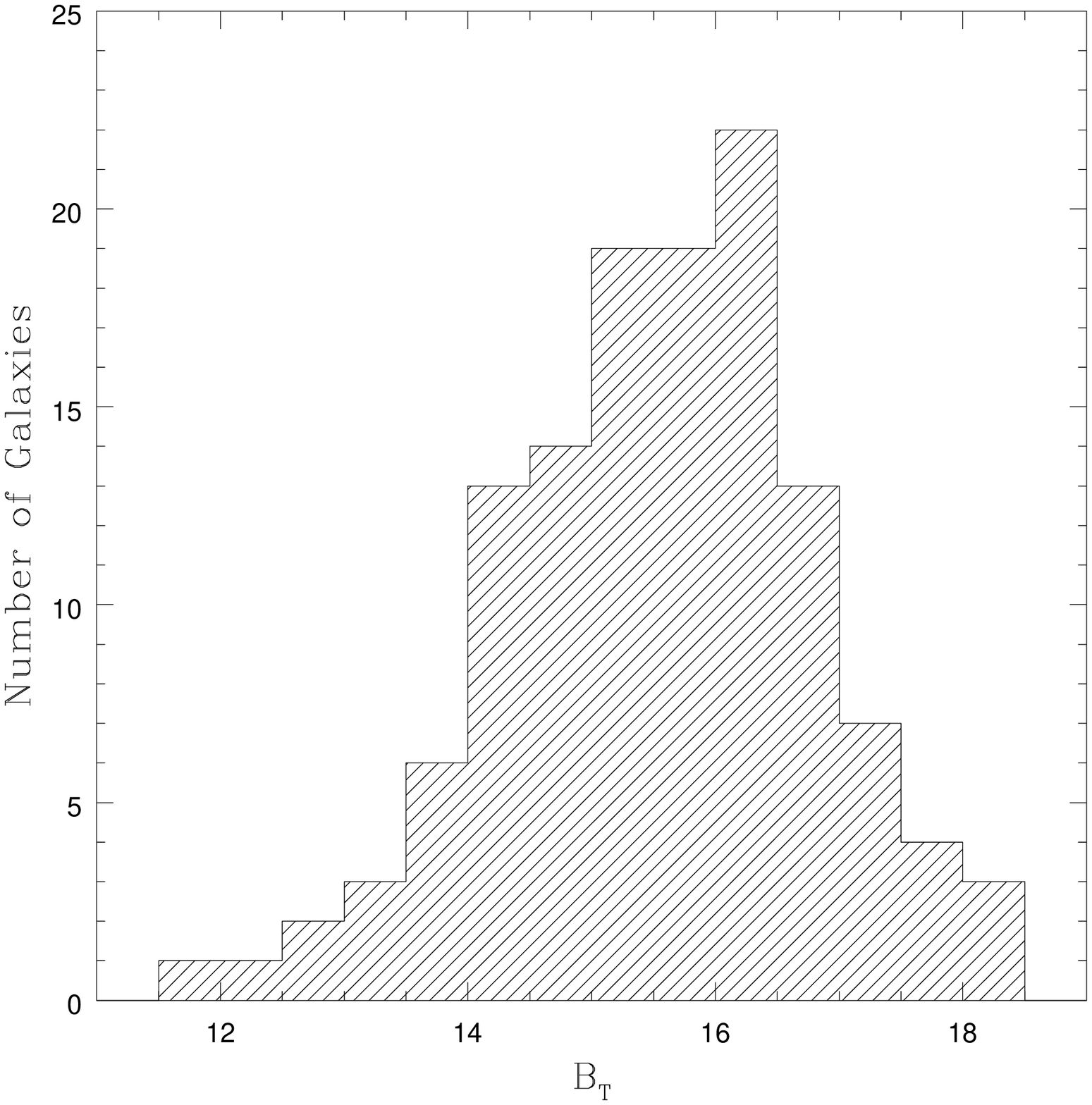,height=90mm,bbllx=70mm,bblly=60mm,bburx=140mm,bbury=250mm}}
\bigskip
\centerline{{\bf Figure 2:} Distribution of $B_T$ magnitude of the 127} \centerline{observed galaxies.}
\end{figure}
\begin{table}[h]
{\bf Table 1:} Observed Hickson Compact Groups (Hickson et al., \cite{Hic:Men}): HCG number, right ascension and declination (1950), number of accordant galaxies, velocity dispersion and mean redshift of group.
\begin{flushleft}
\begin{tabular}{cccccc}
\hline
\hline
\noalign{\smallskip}
HCG & R.A. & Dec. &  $N^{\circ}$ & $\sigma_v$ & z \\
 \ & {\em (1950)} & {\em (1950)} & \ & $(\frac{Km}{s})$ & \ \\ 
\noalign{\smallskip}
\hline
2   &   0 28 48.97   &   8 10 19.3   &   3   &  54.9    & 0.0144 \\
15   &  2 5 2.95     &   1 54 58     &   6   &  426.6   & 0.0228 \\
33   &  5 7 53.69    &   17 57 51    &   4   &  154.9   & 0.026 \\
34   &  5 19 6.72    &   6 38 5.7    &   4   &  316.2   & 0.0307 \\
35   &  8 41 56.87   &   44 42 16.4  &   6   &  316.2   & 0.0542 \\
37   &  9 10 35.78   &   30 12 58    &   5   &  398.1   & 0.0223 \\
38   &  9 24 58.06   &   12 29 58.4  &   3   &  12.9    & 0.0292 \\
43   &  10 8 39.7    &   0 11 32.7   &   5   &  223.9   & 0.033 \\
45   &  10 15 46.72  &   59 21 27.8  &   3   &  182.0   & 0.0732 \\
46   &  10 19 29.69  &   18 6 39.5   &   4   &  323.6   & 0.027 \\
47   &  10 23 7.57   &   13 59 28.2  &   4   &  42.6    & 0.0317 \\
49   &  10 53 19.24  &   67 26 54.2  &   4   &  33.9    & 0.0332 \\
53   &  11 26 18.96  &   21 2 13.9   &   3   &  81.3    & 0.0206\\
54   &  11 26 38.24  &   20 51 38.4  &   4   &  112.2   & 0.0049 \\
56   &  11 29 53.51  &   53 13 16.5  &   5   &  169.8   & 0.027 \\ 
59   &  11 45 53.12  &   12 59 51.3  &   4   &  190.5   & 0.0135 \\
66   &  13 36 47.14  &   57 33 45.5  &   4   &  302     & 0.0699 \\
68   &  13 51 29.15  &   40 33 26.9  &   5   &  154.9   & 0.008 \\
69   &  13 53 12.58  &   25 18 44    &   4   &  223.9   & 0.0294 \\
70   &  14 1 54.07   &   33 34 13.7  &   4   &  144.5   & 0.0636 \\
71   &  14 8 45.02   &   25 44 4.1   &   3   &  416.8   & 0.0301 \\
72   &  14 45 36.94  &   19 16 2.6   &   4   &  263     & 0.0421 \\
74   &  15 17 12.89  &   21 4 31.9   &   5   &  316.2   & 0.0399 \\
75   &  15 19 19.7   &   21 21 45.3  &   6   &  295.1   & 0.0416 \\
76   &  15 29 14.96  &   7 29 20.1   &   7   &  245.5   & 0.034 \\
79   &  15 56 59.93  &   20 53 51    &   4   &  130.0   & 0.0145 \\
81   &  16 15 54.25  &   12 54 57.6  &   4   &  177.8   & 0.0499 \\
82   &  16 26 28.03  &   32 56 21    &   4   &  616.6   & 0.0362 \\
83   &  16 33 12.91  &   6 22 9.6    &   5   &  457.1   & 0.0531 \\
92   &  22 33 40.37  &   33 42 12.6  &   4   &  389.0   & 0.0215 \\
96   &  23 25 28.19  &   8 29 55.4   &   4   &  131.8   & 0.0292 \\
\noalign{\smallskip}
\noalign{\smallskip}
\hline
\hline
\end{tabular} 
\end{flushleft}
\end{table}
\begin{table*}[h]
{\bf Table 2:} Principal features of the observed galaxies (Hickson \cite{Hik}): Galaxy name, heliocentric velocity of group, total photographic blue magnitude, morphological type of galaxy.
\small
\begin{flushleft}
\begin{tabular}{cccc|cccc|cccc}
\hline
\hline
\noalign{\smallskip}
Galaxy & {\em V} & $B_{T}$ & {\em T} & Galaxy & {\em V} & $B_{T}$ & {\em T} & Galaxy & {\em V} & $B_{T}$\ & {\em T} \\
   \   &  $(Km~s^{-1})$ & \ & \       &   \      &  $(Km~s^{-1})$ & \ & \       &   \      &  $(Km~s^{-1})$ & \ & \ \\              
\noalign{\smallskip}
\hline
\noalign{\smallskip}
2a  & 4326  & 13.35 & SBd  & 56a  & 8245  & 15.24 & Sc  & 82a   & 11177 & 14.14 & E3  \\ 
2b  & 4366  & 14.39 & cI   & 56b  & 7919  & 14.5  & SB0 & 82b   & 10447 & 14.62 & SBa \\
2c  & 4235  & 14.15 & SBc  & 56c  & 8110  & 15.37 & S0  & 82c   & 10095 & 14.78 & Im  \\ 
15c & 7222  & 14.37 & E0   & 56d  & 8346  & 16.52 & S0  & 82d   & 11685 & 15.95 & S0a \\
15d & 6244  & 14.65 & E2   & 56e  & 7924  & 16.23 & S0  & 83a   & 15560 & 15.99 & E0  \\
15f & 6242  & 15.74 & Sbc  & 59a  & 4109  & 14.52 & Sa  & 83b   & 16442 & 16.04 & E2  \\      
33a & 7570  & 15.35 & E1   & 59b  & 3908  & 15.2  & E0  & 83c   & 16520 & 16.7  & Scd \\
33b & 8006  & 15.41 & E4   & 59c  & 4347  & 14.4  & Sc  & 83d   & 15500 & 17.91 & Sd  \\
33c & 7823  & 16.4  & Sd   & 59d  & 3866  & 15.8  & Im  & 83e   & 15560 & 18.4  & S0  \\
33d & 7767  & 16.73 & E0   & 66a  & 20688 & 15.38 & E1  & 92b   & 5774  & 13.18 & Sbc \\      
34a & 8997  & 14.2  & E2   & 66b  & 21472 & 16.5  & S0  & 92c   & 6764  & 13.33 & SBa \\
34b & 9620  & 16.56 & Sd   & 66c  & 20801 & 16.39 & S0  & 92d   & 6630  & 13.63 & SB0 \\
34c & 9392  & 16.28 & SBd  & 66d  & 20850 & 17.45 & E2  & 92e   & 6599  & 14.01 & Sa  \\
34d & 8817  & 17.57 & S0   & 68a  & 2162  & 11.84 & S0  & 96a   & 8698  & 13.53 & Sc  \\
35a & 15919 & 15.56 & S0   & 68b  & 2635  & 12.24 & E2  & 96b   & 8616  & 14.49 & E2  \\
35b & 16338 & 15.13 & E1   & 69a  & 8856  & 14.94 & Sc  & 96c   & 8753  & 15.69 & Sa  \\
35c & 16357 & 15.69 & E1   & 69b  & 8707  & 15.59 & SBb & 96d   & 8975  & 16.56 & Im  \\
35d & 15798 & 16.81 & Sb   & 69c  & 8546  & 14.94 & S0  & \     &   \   &   \   & \   \\
35e & 16773 & 17.05 & S0   & 69d  & 9149  & 16.06 & SB0 & \     &   \   &   \   & \   \\
35f & 16330 & 18.12 & E1   & 70d  & 18846 & 15.42 & Sc  & \     &   \   &   \   & \   \\
37a & 6745  & 12.97 & E7   & 70e  & 19117 & 15.91 & Sbc & \     &   \   &   \   & \   \\
37b & 6741  & 14.5  & Sbc  & 70f  & 19243 & 16.4  & SBb & \     &   \   &   \   & \   \\
37c & 7357  & 15.57 & S0a  & 70g  & 19010 & 16.39 & Sa  & \     &   \   &   \   & \   \\
37d & 6207  & 15.87 & Sbdm & 71a  & 9320  & 13.75 & SBc & \     &   \   &   \   & \   \\
37e & 6363  & 16.21 & E0   & 71b  & 9335  & 14.9  & Sb  & \     &   \   &   \   & \   \\
38a & 8760  & 15.25 & Sbc  & 71c  & 8450  & 15.56 & SBc & \     &   \   &   \   & \   \\
38b & 8739  & 14.76 & SBd  & 72a  & 12506 & 13.86 & Sa  & \     &   \   &   \   & \   \\
38c & 8770  & 15.39 & Im   & 72b  & 12356 & 15.48 & S0  & \     &   \   &   \   & \   \\
43a & 10163 & 15.13 & Sb   & 72c  & 13062 & 15.47 & E2  & \     &   \   &   \   & \   \\   
43b & 10087 & 15.18 & SBcd & 72d  & 12558 & 15.64 & SB0 & \     &   \   &   \   & \   \\   
43c & 9916  & 15.82 & SB0  & 74a  & 12255 & 14.06 & E1  & \     &   \   &   \   & \   \\    
43d & 9630  & 16.82 & Sc   & 74b  & 12110 & 15.07 & E3  & \     &   \   &   \   & \   \\    
43e & 9636  & 17.2  & S0   & 74c  & 12266 & 16.1  & S0  & \     &   \   &   \   & \   \\    
45a & 21811 & 15.2  & Sa   & 74d  & 11681 & 16.32 & E2  & \     &   \   &   \   & \   \\   
45b & 22195 & 17.24 & S0a  & 74e  & 11489 & 17.8  & S0  & \     &   \   &   \   & \   \\    
45c & 21799 & 17.6  & Sc   & 75a  & 12538 & 15.2  & E4  & \     &   \   &   \   & \   \\    
46a & 8201  & 16.4  & E3   & 75b  & 12228 & 14.9  & Sb  & \     &   \   &   \   & \   \\   
46b & 8571  & 16.28 & S0   & 75c  & 12292 & 15.93 & S0  & \     &   \   &   \   & \   \\  
46c & 7906  & 16.13 & E1   & 75d  & 12334 & 15.82 & Sd  & \     &   \   &   \   & \   \\    
46d & 7703  & 16.11 & SB0  & 75e  & 12300 & 16.36 & Sa  & \     &   \   &   \   & \   \\    
47a & 9581  & 14.61 & SBb  & 75f  & 13080 & 16.66 & S0  & \     &   \   &   \   & \   \\    
47b & 9487  & 15.67 & E3   & 76a  & 10054 & 15.08 & Sa  & \     &   \   &   \   & \   \\    
47c & 9529  & 16.63 & Sc   & 76b  & 10002 & 14.44 & E2  & \     &   \   &   \   & \   \\    
47d & 9471  & 16.2  & Sd   & 76c  & 10663 & 14.73 & E0  & \     &   \   &   \   & \   \\    
49a & 9939  & 15.87 & Scd  & 76d  & 10150 & 15.21 & E1  & \     &   \   &   \   & \   \\      
49b & 9930  & 16.3  & Sd   & 76e  & 10328 & 16.65 & SB0 & \     &   \   &   \   & \   \\    
49c & 9926  & 17.18 & Im   & 76f  & 10216 & 16.48 & Sc  & \     &   \   &   \   & \   \\   
49d & 10010 & 16.99 & E5   & 79a  & 4294  & 14.35 & E0  & \     &   \   &   \   & \   \\   
53a & 6261  & 12.91 & SBbc & 79b  & 4446  & 13.78 & S0  & \     &   \   &   \   & \   \\  
53b & 6166  & 14.73 & S0   & 79c  & 4146  & 14.72 & S0  & \     &   \   &   \   & \   \\ 
53c & 6060  & 14.81 & SBs  & 79d  & 4503  & 15.87 & Sdm & \     &   \   &   \   & \   \\ 
54a & 1397  & 13.86 & Sdm  & 81a  & 14676 & 16.25 & Sc  & \     &   \   &   \   & \   \\   
54b & 1412  & 16.08 & Im   & 81b  & 15150 & 16.51 & S0  & \     &   \   &   \   & \   \\   
54c & 1420  & 16.8  & Im   & 81c  & 15050 & 17.18 & S0  & \     &   \   &   \   & \   \\ 
54d & 1670  & 18.02 & Im   & 81d  & 14954 & 17.14 & S0a & \     &   \   &   \   & \   \\ 
\noalign{\smallskip} 
\hline
\hline  
\end{tabular}   
\end{flushleft}  
\end{table*}  
\normalsize  
  
%%%%%%%%%%%%%%%%%%%%%%%%%%%%%%% OBSERVATION %%%%%%%%%%%%%%%%  
\section{Observations}  
  
The CCD images of the HCG sample were obtained during three  different observing runs (November 1995, April 1996 and February 1997).
  
Observations  have been carried out at the 2.1 meter telescope (design Ritchey-Chretien) at the National Observatory of Mexico in S. Pedro Martir (SPM). The SPM Cassegrain focus (f/7.5) was coupled with a Tektronix CCD of 1024x1024 pixels, each 24$\mu$m x 24$\mu$m. The telescope scale (13 arcsec/mm) and the pixel dimensions provide  a pixel size of 0.3 arcsec/pix with a resulting field of view of 5.12$^\prime\times$5.12$^\prime$. The CCD gain is 4 e$^-$/ADU.\\
During these three runs we observed 31 HCGs.  
All images were obtained with seeing conditions in the range 2-2.6 arcsec.  
For each HCG two CCD images were taken: the {\em on image}, by using a narrow-band interference filter ({\em  H$_\alpha^{on}$ filter}) centered on the wavelength of the H$_\alpha$  line redshifted to the {\em z} of the galaxy 
(which isolates the H$_\alpha$ emission-line and underlying   
continuum), and the {\em off image}, by using another   
interference filter ({\em H$_\alpha^{off}$ filter}) of similar bandwidth but centered on  an adjacent region of the spectrum (isolating  continuum light only).
Table 3 describes the features of narrow band filters used in this work. In the third column the range of recession velocity that a galaxy should have to give out its H$_\alpha$ line through the  interferential filter is shown.\\
\begin{table}[h]
{\bf Table 3:} Features of interferometric filters: central wavelength, FWHM and corresponding velocity interval
\begin{center}
\begin{tabular}{ccr}
\hline
\hline
\noalign{\smallskip}
$\lambda_{central}$ & FWHM  & Velocity Interval\\
{\em (\AA)}  &  {\em (\AA)}  &  $(Km~s^{-1})$ \\
\noalign{\smallskip}
\hline
\noalign{\smallskip}
 6546 & 81 & $-2628 \rightarrow ~ 1074$\\
 6564 & 72 & $-1600 \rightarrow ~ 1691$\\
 6603 & 80 & $0 \rightarrow ~ 3657$\\
 6607 & 89 & $-23 \rightarrow ~ 4045$\\
 6641 & 79 & $1760 \rightarrow ~ 5371$\\
 6643 & 80 & $1828 \rightarrow ~ 5485$\\
 6683 & 80 & $3657 \rightarrow ~ 7314$\\
 6690 & 91 & $3725 \rightarrow ~ 7885$\\
 6723 & 80 & $5485 \rightarrow ~ 9142$\\
 6732 & 74 & $6034 \rightarrow ~ 9416$\\
 6742 & 85 & $6240 \rightarrow 10216$\\
 6819 & 86 & $9736 \rightarrow 13668$\\
 6920 & 88 & $1431 \rightarrow 18330$\\
 7027 & 93 & $1908 \rightarrow 23335$\\ 
\noalign{\smallskip}
\hline
\hline
\end{tabular} 
\end{center}
\end{table}
In order to calibrate our data, we have observed some spectrophotometric stars, equally spaced in time during each night, from the list of Massey $\&$ Strobel (\cite{Mas:Str}). Table 4 lists the  standards used.
The spectrophotometric standards were observed in the same H$_\alpha$ narrow-band interference filters used to observe Hickson Compact Groups.

\begin{table}[h]
{\bf Table 4:} Standard stars used for calibration
\begin{center}
\begin{tabular}{ccc}
\hline
\hline
\noalign{\smallskip}
Star & $\alpha_{1950}$ & $\delta_{1950}$  \\
\    &  ($h$ $m$ $s$)          &  ($\degr$ $\arcmin$ $\arcsec$) \\            \noalign{\smallskip}
\hline
\noalign{\smallskip}
PG0205+134  & 02 05 21.3 & +13 22 18  \\
Hiltner 600 & 06 42 37.2 & +02 11 25  \\
PG0939+262  & 09 39 58.8 & +26 14 42  \\
Feige34     & 10 36 41.2 & +43 21 50  \\
PG1121+145  & 11 21 39.4 & +14 30 18  \\  
Feige66     & 12 34 54.7 & +25 20 31  \\
Feige67     & 12 39 18.9 & +17 47 24  \\
Kopf27      & 17 41 28   & +05 26 04  \\  
\noalign{\smallskip}
\hline
\hline
\end{tabular} 
\end{center}
\end{table}

The flux from [N$_{II}$] emission lines ($\lambda$=6548 {\AA} and $\lambda$=6584 {\AA}) is included in the on observations. Therefore the flux and luminosity here estimated refer to the sum of H$_\alpha$ and [N$_{II}$] emission lines and not only to H$_\alpha$. Nevertheless through this paper we refer for simplicity to them as  f$_{H_\alpha}$ and L$_{H_\alpha}$ respectively.
The aim of the observations was to study the recent star formation rate occurring in HCG galaxies.
Since the $H_\alpha+[N_{II}$] emission a good star formation tracer as well as the H$_\alpha$ line alone (Kennicutt  \&  Ken, \cite{ken:Ken} ),  the presence of [N$_{II}$] does not invalidate our data. Nevertheless, since the H$_\alpha$/[N$_{II}$] ratio is not constant with radius in the largest galaxies, we will refer to the  global star formation rate of galaxies, that is to the rate
 integrated over all the emitting area of each galaxy.
\vskip 0.2truecm
In Table 5  the journal of the observations is reported as follows:\\
Col.1: Name of the groups;\\
Col.2: Observing date (mm-yy);\\
Col.3: Central wavelength for the {\em H$_\alpha^{on}$ filter} used (\AA);\\
Col.4: Integration time for {\em H$_\alpha^{on}$ filter} exposure (s);\\
Col.5: Central wavelength for the {\em H$_\alpha^{off}$ filter} used (\AA);\\
Col.6: Integration time for {\em H$_\alpha^{off}$ filter} exposure (s).

\begin{table}[h]
{\bf Table 5:} Journal of observations
\begin{center}
\begin{tabular}{cccccc}
\hline
\hline
\noalign{\smallskip}
Group &   Date     &  H$_\alpha^{on}$  &  T$_{exp}$  & H$_\alpha^{off}$  &  T$_{exp}$ \\
\      & {\em (mm-yy)}      & {\em \AA}                      &  {\em (s)}      & {\em \AA}            &  {\em (s)} \\
\noalign{\smallskip}
\hline
\noalign{\smallskip}
HCG2  & Nov.95  & 6643 & 1800 & 6723 & 1800 \\
HCG15 & Nov.95  & 6723 & 1800 & 6643 & 1800 \\
HCG33 & Nov.95  & 6723 & 1800 & 6643 & 1800 \\
HCG34 & Feb.97  & 6732 & 1800 & 6564 & 1800 \\
HCG35 & Feb.97  & 6920 & 1800 & 6690 & 1800 \\
HCG37 & Nov.95  & 6723 & 1800 & 6643 & 1800 \\
HCG38 & Nov.95  & 6723 & 1800 & 6643 & 1800 \\
HCG43 & Feb.97  & 6819 & 1800 & 6607 & 1800 \\
HCG45 & Feb.97  & 7027 & 1800 & 6819 & 1800 \\
HCG46 & Nov.95  & 6723 & 1800 & 6643 & 1800 \\
HCG47 & Apr.96  & 6732 & 1800 & 6564 & 1800 \\
HCG49 & Apr.96  & 6819 & 1800 & 6690 & 1800 \\
HCG53 & Apr.96  & 6690 & 1800 & 6607 & 1800 \\
HCG54 & Nov.95  & 6603 & 1200 & 6683 & 1200 \\
HCG56 & Apr.96  & 6732 & 1800 & 6564 & 1800 \\
HCG59 & Apr.96  & 6690 & 1800 & 6607 & 1800 \\
HCG66 & Apr.96  & 7027 & 1800 & 6920 & 1800 \\
HCG68 & Feb.97  & 6607 & 1200 & 6819 & 1200 \\
HCG69 & Apr.96  & 6732 & 1800 & 6564 & 1800 \\
HCG70 & Apr.96  & 7027 & 1800 & 6920 & 1800 \\
HCG71 & Feb.97  & 6732 & 1800 & 6564 & 1800 \\
HCG72 & Apr.96  & 6819 & 1800 & 6690 & 1800 \\
HCG74 & Feb.97  & 6819 & 1800 & 6607 & 1800 \\
HCG75 & Apr.96  & 6819 & 1800 & 6690 & 1800 \\
HCG76 & Apr.96  & 6819 & 1800 & 6690 & 1800 \\
HCG79 & Apr.96  & 6690 & 1800 & 6607 & 1800 \\
HCG81 & Apr.96  & 6920 & 1800 & 6819 & 1800 \\
HCG82 & Apr.96  & 6819 & 1800 & 6920 & 1800 \\
HCG83 & Apr.96  & 6920 & 1800 & 6819 & 1800 \\
HCG92 & Nov.95  & 6723 & 1800 & 6643 & 1800 \\
HCG96 & Nov.95  & 6723 & 1800 & 6643 & 1800 \\
\noalign{\smallskip}
\hline
\hline
\end{tabular} 
\end{center}
\end{table}

%%%%%%%%%%%%%%%%%%%%%%%%%%%%%%% DATA REDUCTION %%%%%%%%%%%%%%%%
\section {Data Reduction}

\subsection{Bias and Flat-Field Corrections}
The science frames have been first bias subtracted.
For each observing run, we have obtained the proper bias by combining  several bias frames with a median filter.
Then the images have been corrected for pixel to pixel response variations.
For each night, its own flat field has been constructed by medianing several flat field frames carried out during the night.\\
Two different types of flat fields have been used during the three 
observing runs:
the first one has been obtained by medianing twilight sky images and it has been used to reduce the data of 1996. 
The other one, used in 1995 and 1997 runs, has been constructed by medianing flat field frames taken on the dome illuminated with twilight sky. No significant differences have been measured by using the two flat fields.\\
These two steps of data reduction are based on the NOAO IRAF package, developed at the Center for Astrophysics.\\
Finally, cosmic rays and bad pixels have been removed from each frame using Munich Image Data Analysis System (MIDAS).\\

\subsection{ H$_\alpha$ Emission-Line Map} 
The map of the H$_\alpha$ emission-line flux for each HCG  ({\em H$_\alpha$ image}) has been obtained by removing the contribution of the underlying 
continuum, that it is by subtracting the H$_\alpha^{off}$
 from the H$_\alpha^{on}$ (Pogge, \cite{Pog}).
There are several reasons why  the number of 
continuum photons per integration time unit passing through the {\em on filter}
can be different from the one through the {\em off filter}.
For example differences between the transmission curves of the two
narrow-band filters, such as different width and/or
transmission peak; or variations of the sky transparency during the night.
This implies that in order to obtain the H$_\alpha$ emission-line flux
image a careful estimation of the underlying continuum to subtract
from the H$_\alpha^{on}$ is required.
In practice, the H$_\alpha^{off}$ have to be rescaled to the continuum of
the H$_\alpha^{on}$ wavelength.
Since stars do not show H$_\alpha$ emission, the number of continuum 
photons coming from the stars in each HCG field and passing through the 
$on$ and $off$ filters have to be the same.
In-fact, although the $on$ and $off$ filters are sometimes separated by more than 150\AA,  
implying that the number of photons coming from the stars is not rigorously the same, 
such a difference is negligible.
Thus for each HCG field (and for each couple of filters) we have selected at least three
stars and we have calculated the mean scaling factor $K$

\begin{equation}
K=\frac{1}{N}\sum_{i=1}^{N} \left(\frac{C_{on}}{C_{off}}  \right)_i=\left<\left(\frac{C_{on}}{C_{off}} \right) \right>
\end{equation}
where $C_{on}$ and $C_{off}$ are the counts from stars in the $on$ and $off$
image respectively, and N is the number of stars.
Once rescaled, the H$_\alpha^{off}$ have been spatially aligned to the
H$_\alpha^{on}$.
The alignment has been performed by applying the IRAF tasks {\em geomap/geotran}
using the position of at least five stars in the field as 
reference coordinates.
Finally, after having additively rescaled  the $on$ and $off$ images
to the same median value, we have subtracted the H$_\alpha^{off}$
from the H$_\alpha^{on}$ thus obtaining  the image of the H$\alpha$
emission-line flux.

\section{Photometric Calibration}
We have measured instrumental magnitudes  of 
standard stars by constructing for each star its growth curve
through circular concentric apertures.
The magnitude has been taken at the convergence of the curve.
From the spectral energy distributions of our observed  standard stars 
(Massey \& Strobel 1988), we have derived their apparent magnitudes at the
effective wavelengths $\lambda_{eff}$ of our filters, through the
usual relation
\begin{equation}
m_{vF}=-2.5 \cdot \log_{10}f_{\lambda_{eff}}(m_{vF})+2.5\log_{10}f_{\lambda_{eff}}(0) 
\end{equation}
where F is a generic filter, $f_{\lambda_{eff}}(m_{vF})$ and $f_{\lambda_{eff}}(0)$ 
are the spectral irradiances in erg cm$^{-2}$ s$^{-1}$ \AA$^{-1}$ within the 
$F$ filter having the effective wavelength $\lambda_{eff}(F)$
 of a star of magnitude m$_{vF}$ and of a star of $m_{vF}$=0 
 respectively.
From each star we have derived the zero point $Z_p$ of the photometric
calibration for the different filters and nights.

The standard deviation of the zero point values thus obtained is within 0.05
mag during all but one night.
During this night the scatter is much larger than a factor of four.
Thus with the aim at maintaining the uncertainty on the galaxy photometry
within few hundreds percent, we have considered only those galaxies
observed during photometric nights, i.e. those nights for which $\sigma_{Z_p}\le0.05$ mag.

\section{The H$_\alpha$ Emission of Galaxies}
\subsection{Instrumental H$_\alpha$ Fluxes}
Following the data reduction steps  described in section \S 4, we have 
obtained 31 emission-line images, one for each HCG of our sample. 
We have  computed both  isophotal and  adaptive-aperture H$_\alpha$ 
 fluxes  for the HCG galaxies in the 31 fields by  using  SExtractor 
(Bertin et al. \cite{Ber}).
The full analysis of each image is divided in six steps: 
sky background estimation, thresholding, deblending, filtering of the detections, photometry and star/galaxy separation.
For each continuum-subtracted H$_\alpha$ image we have used a detection
 threshold of one sigma above the background.
The H$_\alpha$ isophotal  fluxes have been computed within the region
defined by  the detection threshold.
In addition to the isophotal flux we have  also considered the  corrected 
isophotal  flux estimated by SExtractor that should take into account
  the fraction of flux lost by the isophotal one  (Bertin et al. \cite{Ber}).
In addition  the adaptive-aperture photometry has also been calculated
 (Kron {\cite{Kro:Kro}; Bertin et al. 1996).\\
Out of the 127 accordant observed galaxies belonging to the 31 HCG of our sample, we have been able 
to compute  isophotal  and  adaptive-aperture photometry
for 73 and  69 galaxies respectively.
The $1\sigma$ limiting flux, integrated within the mean seeing disk  (2.3 arcsec), reached in our observations ranges between $1.43 \cdot 10^{-16}$ and $4.13 \cdot 10^{-17}$ $erg ~ cm^{-2} s^{-1}$.\\ 
For 22 galaxies, which have not been detected in our {\em H$_\alpha$} images, 
we have computed the 3$\sigma$ upper limits  above the background:
\begin{equation}
f_{ul}= 3 \cdot rms \cdot \left[\left(\frac{FWHM}{2} \right)^2 \cdot \pi \right]^{\frac{1}{2}}
\end{equation}
being
\[
      \begin{array}{lp{0.6\linewidth}}
	rms &  the  sky estimation accuracy $(counts~pix^{-1} s^{-1})$;\\

	\left[\left(\frac{FWHM}{2} \right)^2 \cdot \pi \right]^{\frac{1}{2}} &  the squareroot of the seeing area in pixels.\\
      \end{array}
   \]

For the remaining 32 galaxies we have not been able to estimate
 the  H$_\alpha$ fluxes because of one of the following reasons: 
\begin{enumerate}
\item the night was not photometric ($\sigma_{Z_p}>>0.05$ mag);
\item the proper narrow band interference filter was not available;
\item too much imperfections are present on the {\em H$_\alpha$} image probably
due to large variations in  seeing conditions between the {\em on} and 
{\em off} band exposures, or due to changes in the telescope focus (e.g. because
of  substantially different thickness of the filters and/or temperature
variations).
\end{enumerate}

In Table 6 we list  the galaxies for which it was not possible 
to measure their flux and the corresponding reason (1,2,3).

\subsection{Zero Point Correction}
In order to obtain  calibrated fluxes and  luminosities for our sample of galaxies, 
we have estimated the zero point flux correction, Z$_{flux}$ such that
\begin{equation}
\label{eq:3.13}
f_{H_\alpha}=(f_{on}-f_{off})=Z_{flux} \cdot \left[C_{on}-{C}_{offn} \right]
\end{equation}
where:
\[
      \begin{array}{lp{0.7\linewidth}}
	f_{on}, f_{off} & are  the isophotal or aperture H$_\alpha$ fluxes of galaxy in 
	the {\em on} and the scaled {\em off filters} respectively;\\
	C_{on}, C_{off} & are  the  counts of the galaxy in the {\em on} and {\em off} band images respectively;\\

      \end{array}
   \]
It can be proved that  the Z$_{flux}$ coefficient of each galaxy is proportional 
to Z$_{flux_{on}}$ i.e. the zero point flux correction of the {\em on} band
image.
Knowing the  Z$_{flux_{on}}$ in magnitudes ($Z_{p_{on}}$, see \S 5) and the 
extinction 
coefficient $k_{on}$ of the site relative to each filter, we have derived
the correction factor Z$_{flux_{on}}$ as follows:
\begin{equation}
Z_{flux_{on}}=\Delta\lambda \cdot 10^{-0.4(Z_{p_{on}}-(k_{on} \cdot X_{s})-b)}
\end{equation}
where $X_s$ is the airmass of the standard star and  $b$ is 
\begin{equation}
b=2.5 \cdot \log_{10}f_{\lambda_{eff}}(0)
\end{equation}
Thus $Z_{flux}$ is given by
\begin{equation}
Z_{flux}=\frac{Z_{flux_{on}}}{10^{[-0.4(k_{on} \cdot X_{s})]}} \cdot 10^{[0.4(k_{on} \cdot X_{g})]}
\end{equation}
where $X_g$ is the airmass of the target galaxy.
This zero point correction was applied to the H$_\alpha$ instrumental
 fluxes and to the upper limits estimated for the undetected  galaxies.
Fluxes and upper limits have been also corrected so that the H$_\alpha$ 
emission-line of the galaxy passes exactly in the center of the corresponding 
{\em on}  filter band, i.e. for the percentage of total flux lost if
 the H$_\alpha$ emission line of the galaxy does not pass exactly in 
the center of the corresponding {\em on} filter.
The corrected fluxes are reported in Tables 7, 8 and 9.

\subsection{Galactic and Internal Extinction Correction}

The H$_\alpha$ fluxes  have  been then corrected  for the galactic 
extinction due to the  gas and the dust of our Galaxy. 
For each target galaxy we have computed  the relative galactic hydrogen 
column density N$_h$ ({\em atoms cm$^{-2}$}) as a function of the galaxy  
coordinates (R.A. and  Dec.).
N$_h$ was obtained interpolating the data available from the 
Stark et al. (\cite{Sta:Sta}) data-base.
We computed also an interpolation error defined as the mean of 
differences weighted on the distances.
Using the relations:
\begin{equation}
\frac{N_{h}}{A_{B}-A_{V}}= \frac{N_{h}}{E(B-V)}=5.2 \cdot 10^{21}\ atoms\ cm^{-2}\ mag^{-1}
\end{equation}
and:
\begin{equation}
R=\frac{A_{V}}{E(B-V)}= 3.1
\end{equation}
(Ryder \& Dopita \cite{Ryd:Dop}) we have derived the visual extinction 
coefficients A$_V$ and A$_B$ ({\em mag}) for each galaxy.
Following Ryder \& Dopita (\cite{Ryd:Dop}), we have obtained the 
multiplicative correction $\alpha_G$ to apply to the  H$_\alpha$ flux:
\begin{equation}
\alpha_G= 10^{(0.4 \cdot A_{H_{\alpha}})}= 10^{(0.4 \cdot 0.64A_{B})}
\end{equation}
where $A_{H_{\alpha}}$ is the H$_\alpha$ extinction coefficient in magnitudes.
 The isophotal fluxes corrected for Galactic Extinction are reported in Table 7.
The minimum and maximum values obtained for $\alpha_G$ are respectively 1.04 and 2.68.\\
The H$_\alpha$ fluxes of spirals  have also been  corrected  for the 
Internal Extinction due to the interstellar medium inside the target galaxy
itself. 
This correction in the  blue band is usually obtained by summing to the galaxy
magnitude the value
\begin{equation}
A_i=c_B \cdot log (r_i)
\end{equation}
(Haynes \& Giovanelli \cite{Hay:Gio})
where:
\[
      \begin{array}{lp{0.9\linewidth}}
	c_B & is a morphological type dependent correction coefficient in  
 the $B$ band and \\
	r_i & is the intrinsic axial ratio of the galaxy.\\
      \end{array}
   \]
On the basis of the interstellar extinction curve (e.g. Osterbrook 1974)
we have derived the H$_\alpha$ extinction correction term $c_{H_\alpha}$
using the following transformation:
\begin{equation}
c_B \cdot log(r_i) - c_{H_\alpha} \cdot log(r_i) = -2.5 \cdot log(e_B) + 2.5 \cdot log(e_{H_\alpha})
\end{equation}
where e$_B$ and e$_{H_\alpha}$  are  the extinction values at the effective  
wavelength respectively of the $B$ and the H$_\alpha$ filters .
Finally we have obtained the flux correction factor $\alpha_i=10^{[0.4 \cdot c_{H_\alpha} \cdot Log(r_i)]}$, to apply 
 to  our spiral galaxies. The minimum and maximum values obtained for $\alpha_i$ are respectively 1.1 and 1.6.\\

On the basis of the fluxes thus obtained we have derived the H$_\alpha$ 
luminosity L$_{H_{\alpha}}$ of galaxies:
\begin{equation}
L_{H_{\alpha}}=4\pi \cdot f_{H_{\alpha}} \cdot d_{L}^2
\end{equation}
where the luminosity distance $d_L$ is defined as:
\begin{equation}
d_{L}=\frac{c}{H_0 \cdot q_{0}^2} \cdot \left(q_0 \cdot z+(q_0-1) \cdot [-1+(2q_0 \cdot z+1)^{\frac{1}{2}} \right)
\end{equation}
We adopted $H_0=100~km~s^{-1} Mpc^{-1}$ and $q_0=0.5$.

In Table 7 we report the isophotal luminosities of the galaxies ($L_{iso} (1)$) uncorrected for Galactic and Internal Extinction. 

\section{Error Estimate}
The uncertainties reported in Tables 7, 8 and 9 regarding the different
flux estimates have been derived as follows:
\scriptsize
\begin{equation}
\sigma_{f_{H_{\alpha}}}= \sqrt{\left(\frac{\partial f_{H_{\alpha}}}{\partial C_{H\alpha}
} \cdot \sigma_{C_{H\alpha}} \right)^2 + \left(\frac{\partial f_{H_{\alpha}}}{\partial Z
_{flux}} \cdot \sigma_{Z_{flux}} \right)^2 + \left(\frac{\partial f_{H_{\alpha}}
}{\partial \alpha_G} \cdot \sigma_{\alpha_G} \right)^2
} 
\end{equation}
\normalsize
where:
\[
      \begin{array}{lp{0.6\linewidth}}
	C_{H\alpha}=C_{on}-C_{off} & is the H$_\alpha$ instrumental
flux;   \\

	\sigma_{C_{H\alpha}} &  is the uncertainty on $C_{H\alpha}$. 
It is the standard deviation of the stellar flux residuals measured on the 
net H$_\alpha$ images. 
Since stars do not show H$_\alpha$ emission  we should not
detect any flux at the star positions on the net images.
The detection of net counts  could be thus represent simple poisson noise
and/or no perfect continuum subtraction. Therefore the standard deviation 
of such measurements gives a good estimation of the pure and not 
statistical uncertainties on $C_{H\alpha}$; \\

	\sigma_{Z_{flux}} & represents the accuracy on the scale factor
$Z_{flux}$ (\S 6.2) and is given by the difference of the zero points derived by the
two standard stars observed before and after the target HCG;\\ 
	\sigma_{\alpha_G} & is the error about the Galactic 
Extinction $\alpha_G$ (\S 6.3) derived by the propagation error formula
to $\alpha_G$.\\
      \end{array}
   \]

The errors regarding the H$_\alpha$ luminosity (Tables 7 and 9) have been 
calculated by applying the error propagation  formula.

\section{The H$_\alpha$ Catalogue} 

Table 7 lists the H$_\alpha$ isophotal fluxes and luminosities
 estimated:
Col.1: Name of the galaxy (Hickson 1982);\\
Col.2: Flux uncorrected for Galactic and Internal Extinction: $f_{iso}$ (1) ($erg~cm^{-2}~s^{-1}$);\\
Col.3: Error about $f_{iso}$ (1): $\sigma_{f_{iso}}(1)$ ($erg~cm^{-2}~s^{-1}$);\\
Col.4: Luminosity uncorrected for Galactic and Internal Extinction: $L_{iso}$ (1) ($erg~s^{-1}$);\\
Col.5: Error about $L_{iso}$ (1): $\sigma_{L_{1}}$ (1) ($erg~s^{-1}$);\\ 
Col.6: Flux corrected for Galactic Extinction $f_{iso}$ (2) ($erg~cm^{-2}~s^{-1}$) ;\\
Col.7: Error about $f_{iso}$ (2): $\sigma_{f_{iso}}$ (2) ($erg~cm^{-2}~s^{-1}$);\\
Col.8: Flux corrected for  Galactic and Internal Extinction for spiral galaxies: $f_{iso}$ (3) ($erg~s^{-1}$);\\
Col.9: Isophotal area at 1$\sigma$ above the background: $A_{iso}$ ($arcsec^2$);\\
Col.10: {\em S/N} ratio computed within the isophotal region defined by the detection threshold.
\vskip 0.5truecm
In Tables 8 and 9 the fluxes are not corrected for Galactic and Internal Extinction. Such corrections can be simply derived from Table 7.
\vskip 0.5truecm

Table 8 lists isophotal corrected and adaptive aperture fluxes:\\
Col.1: Name of the galaxy (Hickson 1982);\\
Col.2: Isophotal flux $f_{isocor}$ ($erg~cm^{-2}~s^{-1}$);\\
Col.3: Error about $f_{isocor}$: $\sigma_{isocor}$ ($erg~cm^{-2}~s^{-1}$);\\
Col.4: Kron flux $f_{Kron}$ ($erg~cm^{-2}~s{-1}$);\\
Col.5: Error about $f_{Kron}$: $\sigma_{Kron}$ ($erg~cm^{-2}~s{-1}$).\\
\vskip 0.5truecm
 
In Table 9 we report the fluxes and luminosities of upper limits:\\
Col.1: Name of the galaxy (Hickson 1982);\\
Col.2: Flux at 3 $\sigma$ above the background ($f_{3\sigma}$) ($erg~cm^{-2}~s^{-1}$);\\
Col.3: Error about $f_{3\sigma}$: $\sigma$ ($erg~cm^{-2}~s^{-1}$);\\
Col.4: Luminosity at 3 $\sigma$ above the background ($L_{3\sigma}$) ($erg~cm^{-2}~s^{-1}$);\\
Col.5: Error about $L_{3\sigma}$: $\sigma$ ($erg~cm^{-2}~s{-1}$).\\
\vskip 0.5truecm

We have compared the estimated $f_{iso}$, $f_{isocor}$ and $f_{kron}$ for the detected galaxies by using the Kolmogorov-Smirnov test: we found they are drawn from the same parent population. The  distributions of the three different fluxes estimated for the detected galaxies are shown in Figure 3. 
In the following we make use of $f_{iso}$ in our considerations unless it is differently specified.
\begin{figure}[h]
\centerline{\psfig{figure=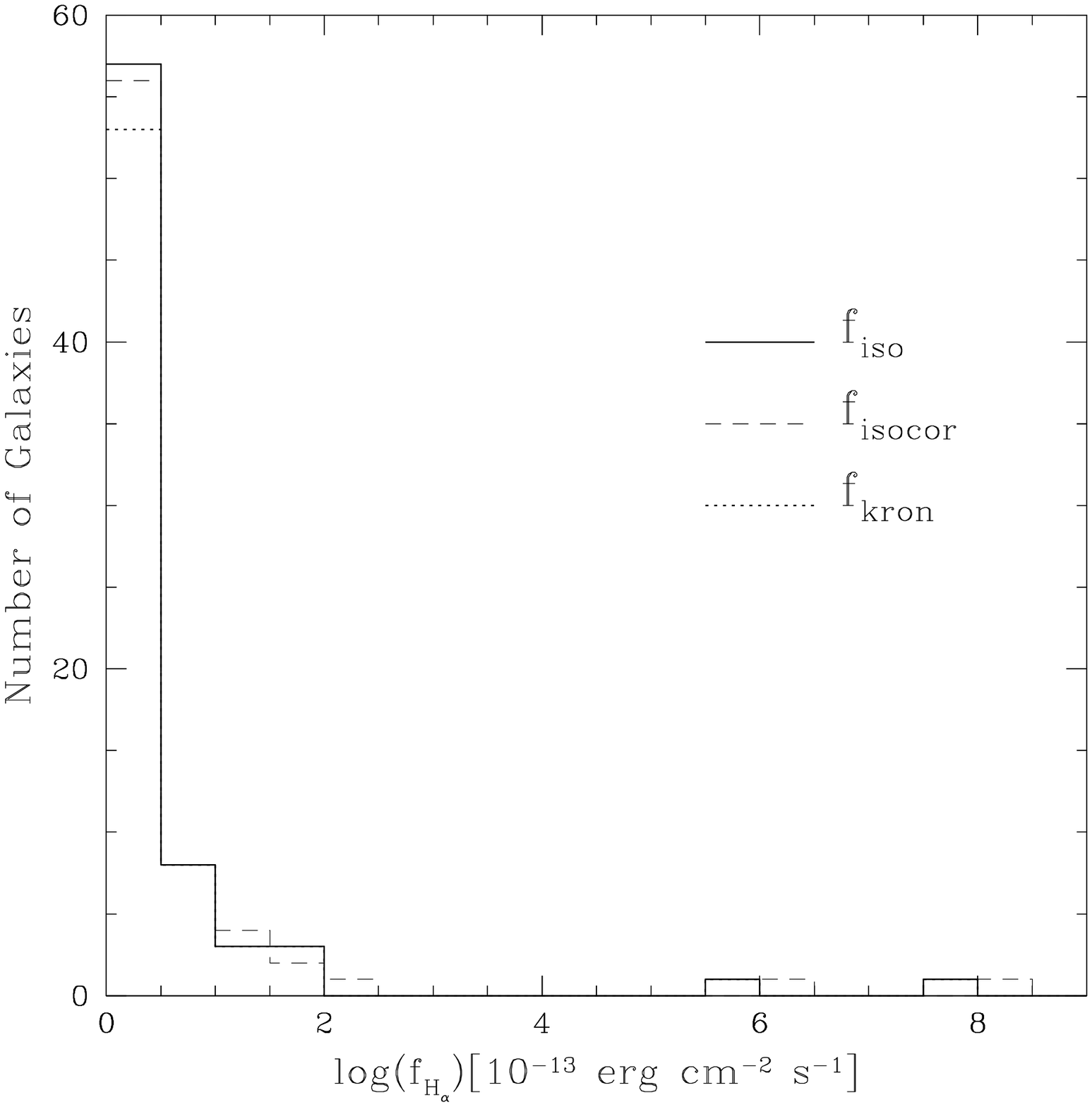,height=80mm,bbllx=70mm,bblly=60mm,bburx=140mm,bbury=250mm}}
\bigskip
\bigskip
{{\bf Figure 3:} Distribution of $f_{iso}$ (solid line), $f_{isocor}$ (dashed line) and $f_{kron}$ (dotted line) (corrected for Galactic Extinction) of the 73 detected galaxies. The width of each bin is 0.5 [$10^{-13}$ erg cm$^{-2}$ s$^{-1}$].}
\end{figure}
\vskip 0.5truecm
In Figure 4 the distribution of $H_{\alpha}$ isophotal luminosity $L_{iso}$, corrected for Galactic Extinction and uncorrected for Internal Extinction, of the 73 detected galaxies (shaded histogram) is shown. We over-plot also the  distribution of $H_{\alpha}$ upper limits to luminosity for the 22 undetected galaxies (dashed line).  
\begin{figure}[h]
\centerline{\psfig{figure=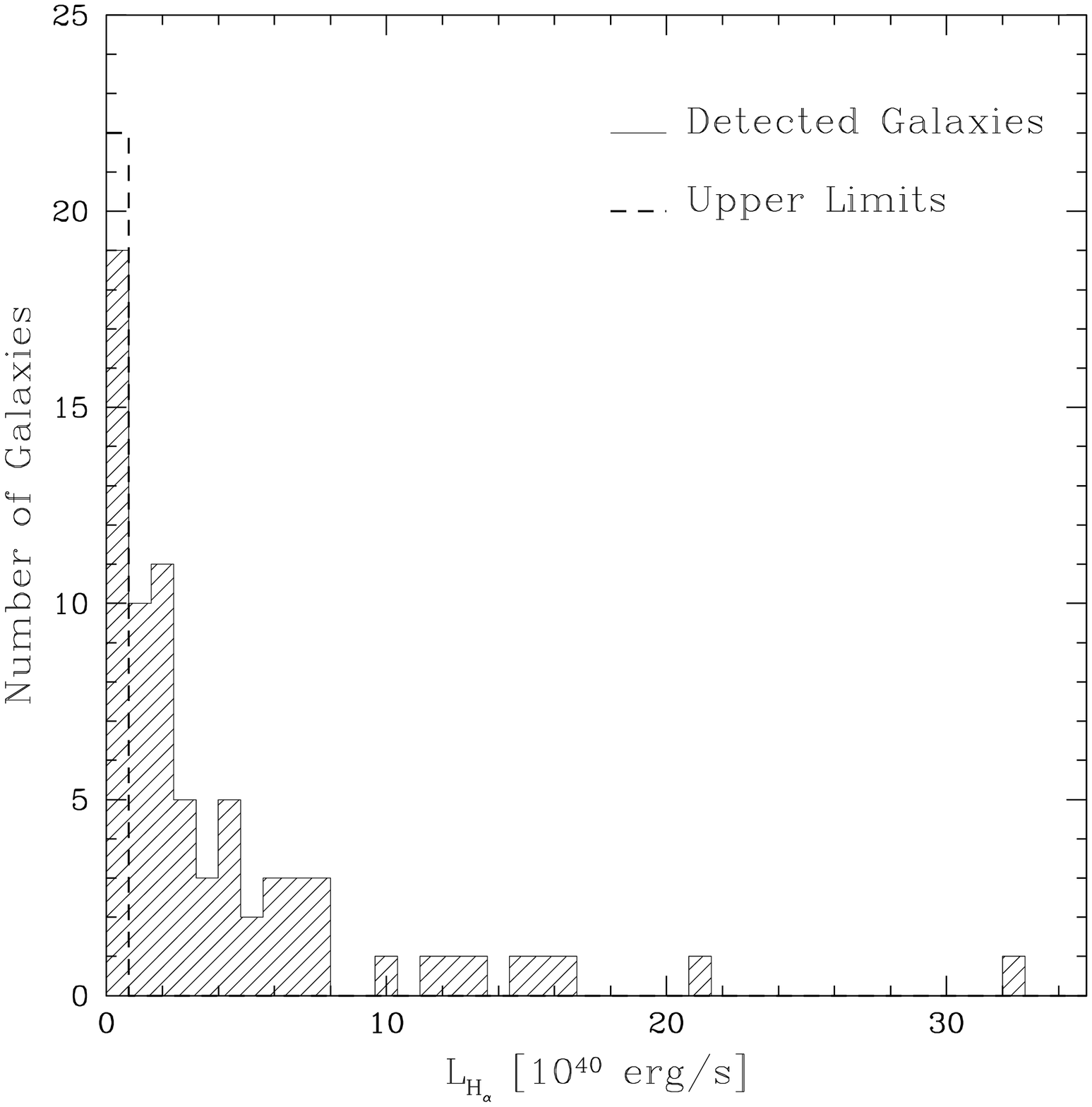,height=80mm,bbllx=70mm,bblly=60mm,bburx=140mm,bbury=250mm}}
\bigskip
\bigskip
{{\bf Figure 4:} Distribution of $H_{\alpha}$  luminosity, corrected for Galactic Extinction, for both the 73 detected galaxies (shaded histogram) and for the 22 upper limits (dashed line). The width of each bin is 0.8 [$10^{40}$ erg s$^{-1}$].}
\end{figure}
\vskip 0.5truecm

\begin{table}[h]
{\bf Table 6:} Galaxies without estimated flux (see \S 6.1)
\begin{center}
\begin{tabular}{cccccc}
\hline
\hline
\noalign{\smallskip}
Galaxy & Reason (1,2,3) & \ & \ & Galaxy & Reason (1,2,3) \\
\noalign{\smallskip} 
\hline
\noalign{\smallskip}
2c  & 1 & \ & \ & 59b & 3 \\ 
15c & 1 & \ & \ & 59c & 3 \\ 
15d & 1 & \ & \ & 59d & 3 \\ 
15f & 1 & \ & \ & 70d & 2 \\ 
34b & 2 & \ & \ & 79a & 3 \\ 
43d & 2 & \ & \ & 79b & 3 \\ 
43e & 2 & \ & \ & 79c & 3 \\ 
47a & 2 & \ & \ & 79d & 3 \\ 
47b & 2 & \ & \ & 81a & 3 \\ 
47c & 2 & \ & \ & 81b & 3 \\
47d & 2 & \ & \ & 81c & 3 \\ 
54a & 3 & \ & \ & 81d & 3 \\ 
54b & 3 & \ & \ & 96a & 1 \\ 
54c & 3 & \ & \ & 96b & 1 \\ 
54d & 3 & \ & \ & 96c & 1 \\ 
59a & 3 & \ & \ & 96d & 1 \\ 
\noalign{\smallskip}
\noalign{\smallskip}
\hline
\hline
\end{tabular} 
\end{center}
\end{table}

\begin{table*}[h]
{\bf Table 7:} Isophotal Fluxes and Luminosities
\normalsize
\begin{flushleft}
\begin{tabular}{cccccccccccc}
\hline
\hline
\noalign{\smallskip}
\noalign{\smallskip}
Galaxy & $f_{iso}$ (1) & $\sigma_{f_{iso}}$ (1) & $L_{iso}$ (1) & $\sigma_{L_{iso}}$ (1) & $f_{iso}$ (2) & $\sigma_{f_{iso}}$ (2) & $f_{iso}$ (3) & $A_{iso}$ & {\em S/N}\\
\noalign{\smallskip} 
\noalign{\smallskip}
 \     & $\left(\frac{erg}{cm^{2}s} \right)$ & $\left(\frac{erg}{cm^{2}s} \right)$ & $\left(\frac{erg}{s} \right)$ & $\left(\frac{erg}{s} \right)$ &  $\left(\frac{erg}{cm^{2}s} \right)$ & $\left(\frac{erg}{cm^{2}s} \right)$ & $\left(\frac{erg}{cm^{2}s} \right)$ & $\left(arcsec^2 \right)$ &  \ \\
\noalign{\smallskip}
\noalign{\smallskip}
\hline
\noalign{\smallskip}
2a	&	6.29E-13	&	3E-14	&	1.39E+41	&	6E+39	&	7.60E-13	&	3E-14	&	9.12E-13	& 	1643.6	&    890 \\ 
2b	&	4.94E-13	&	2E-14	&	1.07E+41	&	5E+39	&	5.98E-13	&	3E-14	&	\		&	315.8	&    1584   \\ 
33a	&	5.69E-15	&	9E-16	&	3.73E+39	&	6E+38	&	1.52E-14	&	2E-15	&	\		&	20.3	&    99    \\ 
33b	&	3.16E-15	&	9E-16	&	2.32E+39	&	6E+38	&	8.47E-15	&	2E-15	&	\		&	16.1	&    60   \\ 
33c	&	4.03E-14	&	3E-15	&	2.82E+40	&	2E+39	&	1.08E-13	&	7E-15	&	1.12E-13	&	269.6	&    191   \\ 
33d	&	1.18E-15	&	8E-16	&	8.16E+38	&	6E+38	&	3.17E-15	&	2E-15	&	\		&	8.2	&    32   \\ 
34c	&	3.76E-14	&	1E-14	&	3.81E+40	&	1E+40	&	7.57E-14	&	3E-14	&	9.33E-14	&	74.9	&    68   \\ 
35a	&	1.46E-14	&	2E-15	&	4.29E+40	&	6E+39	&	1.66E-14	&	2E-15	&	\		&	25.6	&   108    \\ 
35b	&	8.15E-15	&	2E-15	&	2.53E+40	&	6E+39	&	9.28E-15	&	2E-15	&	\		&	21.9	&    67   \\ 
35c	&	9.16E-15	&	2E-15	&	2.85E+40	&	6E+39	&	1.04E-14	&	2E-15	&	\		&	25.7	&    69   \\ 
35d	&	8.95E-15	&	2E-15	&	2.59E+40	&	5E+39	&	1.02E-14	&	2E-15	&	1.10E-14	&	35.7	&    59   \\ 
35e	&	2.09E-15	&	2E-15	&	6.84E+39	&	6E+39	&	2.38E-15	&	2E-15	&	\		&	19.4	&    18   \\ 
35f	&	1.19E-15	&	2E-15	&	3.69E+39	&	6E+39	&	1.36E-15	&	2E-15	&	\		&	5.1	&    20   \\ 
37a	&	5.12E-14	&	9E-15	&	2.66E+40	&	5E+39	&	5.65E-14	&	1E-14	&	\		&	235.2	&    328   \\ 
37b	&	2.49E-14	&	9E-15	&	1.29E+40	&	5E+39	&	2.74E-14	&	1E-14	&	3.77E-14	&	217.1	&    166   \\ 
37c	&	5.17E-15	&	9E-15	&	3.20E+39	&	6E+39	&	5.71E-15	&	1E-14	&	\		&	46.7	&    75    \\ 
37d	&	1.74E-14	&	1E-14	&	7.68E+39	&	5E+39	&	1.93E-14	&	1E-14	&	\		&	98.4	&    129   \\ 
37e	&	3.35E-15	&	1E-14	&	1.55E+39	&	5E+39	&	3.69E-15	&	1E-14	&	\		&	28.4	&    51    \\ 
38a	&	3.67E-14	&	5E-15	&	3.23E+40	&	4E+39	&	4.31E-14	&	6E-15	&	4.60E-14	&	156.3	&    147   \\ 
38b	&	4.29E-14	&	5E-15	&	3.76E+40	&	5E+39	&	5.03E-14	&	6E-15	&	6.53E-14	&	150.4	&    175  \\ 
38c	&	2.06E-14	&	4E-15	&	1.82E+40	&	3E+39	&	2.41E-14	&	4E-15	&	\		&	60.0	&    133   \\ 
43a	&	5.34E-14	&	4E-15	&	6.34E+40	&	4E+39	&	6.19E-14	&	4E-15	&	6.32E-14	&	210.6	&    190  \\ 
43b	&	4.79E-14	&	3E-15	&	5.6E+40		&	4E+39	&	5.55E-14	&	4E-15	&	7.01E-14	&	174.4	&    177  \\ 
43c	&	2.36E-14	&	2E-15	&	2.67E+40	&	2E+39	&	2.74E-14	&	2E-15	&	\		&	105.3	&    82  \\ 
45a	&	3.70E-14	&	2E-15	&	2.06E+41	&	1E+40	&	3.85E-14	&	2E-15	&	4.81E-14	&	150.4	&    136   \\ 
45b	&	7.10E-15	&	2E-15	&	4.1E+40		&	1E+40	&	7.39E-15	&	2E-15	&	\		&	26.7	&    60  \\ 
45c	&	2.78E-15	&	2E-15	&	1.55E+40	&	1E+40	&	2.89E-15	&	2E-15	&	3.52E-15	&	14.0	&    32   \\ 
46a	&	4.72E-15	&	7E-16	&	3.63E+39	&	5E+38	&	5.42E-15	&	8E-16	&	\		&	29.7	&    71   \\ 
46b	&	3.08E-15	&	6E-16	&	2.60E+39	&	5E+38	&	3.54E-15	&	7E-16	&	\		&	17.4	&    60   \\ 
46c	&	1.45E-15	&	6E-16	&	1.04E+39	&	4E+38	&	1.66E-15	&	7E-16	&	\		&	15.1	&    31   \\ 
46d	&	2.96E-16	&	6E-16	&	2.01E+38	&	4E+38	&	3.41E-16	&	7E-16	&	\		&	8.0	&    8.9   \\ 
49a	&	5.05E-14	&	2E-15	&	5.74E+40	&	2E+39	&	5.47E-14	&	2E-15	&	8.64E-14	&	72.8	&    162  \\ 
49b	&	1.18E-13	&	4E-15	&	1.34E+41	&	5E+39	&	1.28E-13	&	5E-15	&	1.75E-13	&	96.3	&    330   \\ 
49c	&	2.35E-14	&	2E-15	&	2.67E+40	&	2E+39	&	2.55E-14	&	2E-15	&	\		&	48.9	&    92  \\ 
49d	&	1.21E-14	&	1E-15	&	1.40E+40	&	1E+39	&	1.31E-14	&	1E-15	&	\		&	27.0	&    72   \\ 
53a	&	1.44E-13	&	4E-15	&	6.44E+40	&	2E+39	&	1.55E-13	&	4E-15	&	1.55E-13	&	429.4	&    162   \\ 
53b	&	1.48E-14	&	4E-15	&	6.41E+39	&	2E+39	&	1.59E-14	&	4E-15	&	\		&	38.4	&    56  \\ 
53c	&	9.37E-14	&	4E-15	&	3.93E+40	&	2E+39	&	1.01E-13	&	4E-15	&	1.30E-13	&	155.3	&    177   \\
\noalign{\smallskip}
\hline
\hline
\end{tabular}
\end{flushleft}
\end{table*} 
\newpage
\normalsize 
\begin{table*}[h]
{\bf Table 7:} Continued
\normalsize
\begin{flushleft}
\begin{tabular}{cccccccccc}
\hline
\hline
\noalign{\smallskip}
\noalign{\smallskip}
Galaxy & $f_{iso}$ (1) & $\sigma_{f_{iso}}$ (1) & $L_{iso}$ (1) & $\sigma_{L_{iso}}$ (1) & $f_{iso}$ (2) & $\sigma_{f_{iso}}$ (2) & $f_{iso}$ (3) & $A_{iso}$ &  {\em S/N}\\
\noalign{\smallskip} 
\noalign{\smallskip}
 \     & $\left(\frac{erg}{cm^{2}s} \right)$ & $\left(\frac{erg}{cm^{2}s} \right)$ & $\left(\frac{erg}{s} \right)$ & $\left(\frac{erg}{s} \right)$ &  $\left(\frac{erg}{cm^{2}s} \right)$ & $\left(\frac{erg}{cm^{2}s} \right)$ & $\left(\frac{erg}{cm^{2}s} \right)$ & $\left(arcsec^2 \right)$ &   \ \\
\noalign{\smallskip}
\noalign{\smallskip}
\hline
\noalign{\smallskip}
56a	&	2.15E-14	&	1E-15	&	1.68E+40	&	8E+38	&	2.26E-14	&	1E-15	&	2.65E-14	&	131.7	&    87  \\ 
56b	&	1.58E-13	&	3E-15	&	1.13E+41	&	2E+39	&	1.66E-13	&	3E-15	&	\		&	150.6	&    601   \\ 
56d	&	2.78E-14	&	1E-15	&	2.22E+40	&	9E+38	&	2.93E-14	&	1E-15	&	\		&	68.5	&    156  \\ 
56e	&	7.86E-15	&	9E-16	&	5.65E+39	&	7E+38	&	8.26E-15	&	1E-15	&	\		&	39.5	&    58    \\  
66b	&	5.64E-14	&	4E-15	&	3.04E+41	&	2E+40	&	5.99E-14	&	4E-15	&	\		&	30.7	&    579   \\ 
69a	&	8.13E-15	&	2E-15	&	7.32E+39	&	1E+39	&	8.67E-15	&	2E-15	&	1.05E-14	&	56.6	&    45    \\ 
69b	&	4.43E-14	&	4E-15	&	3.86E+40	&	4E+39	&	4.73E-14	&	4E-15	&	6.75E-14	&	50.4	&    193    \\ 
70e	&	2.22E-14	&	9E-16	&	9.46E+40	&	4E+39	&	2.35E-14	&	1E-15	&	3.71E-14	&	43.7	&    60    \\ 
70g	&	1.14E-14	&	7E-16	&	4.87E+40	&	3E+39	&	1.21E-14	&	7E-16	&	1.92E-14	&	39.8	&    42   \\ 
71a	&	1.45E-13	&	7E-15	&	1.45E+41	&	7E+39	&	1.56E-13	&	7E-15	&	2.31E-13	&	571.7	&    342  \\ 
71b	&	8.71E-15	&	8E-16	&	8.72E+39	&	8E+38	&	9.36E-15	&	8E-16	&	1.04E-14	&	97.7	&    50    \\ 
71c	&	2.62E-14	&	1E-15	&	2.14E+40	&	1E+39	&	2.81E-14	&	1E-15	&	3.94E-14	&	228.8	&    162   \\ 
72a	&	8.12E-15	&	2E-15	&	1.47E+40	&	4E+39	&	9.04E-15	&	2E-15	&	1.19E-14	&	34.7	&    43  \\ 
72b	&	3.81E-15	&	2E-15	&	6.72E+39	&	4E+39	&	4.24E-15	&	2E-15	&	\		&	18.0	&    28   \\ 
72c	&	2.98E-15	&	2E-15	&	5.87E+39	&	4E+39	&	3.32E-15	&	2E-15	&	\		&	10.9	&    28   \\ 
72d	&	3.83E-15	&	2E-15	&	6.97E+39	&	4E+39	&	4.26E-15	&	2E-15	&	\		&	13.2	&    33   \\
74a	&	2.73E-14	&	2E-15	&	4.73E+40	&	3E+39	&	3.27E-14	&	2E-15	& 	\		&	61.4	&  169   \\ 
74b	&	6.56E-15	&	7E-16	&	1.11E+40	&	1E+39	&	7.84E-15	&	9E-16	&	\		&	40.6	&  52 \\ 
74c	&	5.18E-15	&	7E-16	&	8.98E+39	&	1E+39	&	6.19E-15	&	9E-16	&	\		&	22.0	&  53  \\ 
74d	&	5.73E-15	&	7E-16	&	9.01E+39	&	1E+39	&	6.85E-15	&	8E-16	&	\		&	23.0	&  63  \\ 
75a	&	2.96E-14	&	1E-15	&	5.37E+40	&	2E+39	&	3.56E-14	&	1E-15	&	\		&	137.6	&  86  \\ 
75b	&	6.22E-15	&	6E-16	&	1.07E+40	&	1E+39	&	7.48E-15	&	7E-16	&	8.64E-15	&	21.1	&  46  \\ 
75c	&	9.06E-15	&	7E-16	&	1.58E+40	&	1E+39	&	1.09E-14	&	8E-16	&	\		&	54.6	&  42  \\ 
75d	&	2.95E-14	&	1E-15	&	5.19E+40	&	2E+39	&	3.56E-14	&	1E-15	&	4.50E-14	&	96.3	&  148  \\ 
75f	&	1.36E-15	&	4E-16	&	2.69E+39	&	8E+38	&	1.64E-15	&	5E-16	&	\		&	44.1	&  10 \\ 
76a	&	1.20E-14	&	2E-15	&	1.39E+40	&	2E+39	&	1.42E-14	&	2E-15	&	1.69E-14	&	23.3	&  61  \\ 
76b	&	1.17E-14	&	2E-15	&	1.35E+40	&	2E+39	&	1.4E-14		&	2E-15	&	\		&	55.8	&  38  \\ 
76c	&	1.1E-14		&	2E-15	&	1.44E+40	&	2E+39	&	1.31E-14	&	2E-15	&	\		&	57.4	&  33  \\ 
76d	&	1.18E-14	&	2E-15	&	1.4E+40		&	2E+39	&	1.40E-14	&	2E-15	&	\		&	54.6	&  33  \\ 
76f	&	5.60E-15	&	2E-15	&	6.72E+39	&	2E+39	&	6.91E-15	&	2E-15	&	9.37E-15	&	13.1	&  36  \\ 
82b	&	4.17E-15	&	1E-15	&	5.23E+39	&	2E+39	&	4.51E-15	&	2E-15	&	4.49E-15	&	32.5	&  23  \\ 
82c	&	3.38E-14	&	2E-15	&	3.96E+40	&	2E+39	&	3.66E-14	&	2E-15	&	\		&	83.4	&  109  \\ 
83b	&	1.31E-15	&	1E-15	&	4.12E+39	&	3E+39	&	1.71E-15	&	1E-15	&	\		&	21.4	&  8  \\ 
83c	&	3.03E-14	&	3E-15	&	9.59E+40	&	9E+39	&	3.95E-14	&	4E-15	&	4.95E-14	&	69.3	&  104   \\ 
92c	&	5.10E-14	&	3E-15	&	2.67E+40	&	1E+39	&	7.25E-14	&	4E-15	&	1.07E-13	&	120.9	&  164   \\ 
\noalign{\smallskip}
\hline
\hline
\end{tabular} 
\end{flushleft}
\end{table*}
\normalsize

\begin{table*}[h]
{\bf Table 8:} Isophotal Corrected and Kron Fluxes
\small
\begin{flushleft}
\begin{tabular}{ccccc|ccccc}
\hline
\hline
\noalign{\smallskip}
\noalign{\smallskip}
Galaxy & $f_{isocor}$ & $\sigma_{isocor}$ & $f_{Kron}$ & $\sigma_{Kron}$ & Galaxy & $f_{isocor}$ & $\sigma_{isocor}$ & $f_{Kron}$ & $\sigma_{Kron}$\\
\noalign{\smallskip} 
\noalign{\smallskip}
 \     & $\left(\frac{erg}{cm^2s} \right)$ & $\left(\frac{erg}{cm^2s} \right)$ & $\left(\frac{erg}{cm^2s} \right)$ & $\left(\frac{erg}{cm^2s} \right)$ &  \     & $\left(\frac{erg}{cm^2s} \right)$ & $\left(\frac{erg}{cm^2s} \right)$ & $\left(\frac{erg}{cm^2s} \right)$ & $\left(\frac{erg}{cm^2s} \right)$ \\
\noalign{\smallskip}
\noalign{\smallskip}
\hline
\noalign{\smallskip}
2a	&	6.64E-13	&	3E-14	&	6.37E-13        &       3E-14       & 74a     &       2.87E-14        &       2E-15       &       2.45E-14        &       2E-15       \\ 
2b	&	4.99E-13	&	2E-14	&       4.93E-13        &       2E-14       & 74b     &       6.89E-15        &       8E-16       &       6.78E-15        &       8E-16       \\ 
33a	&	6.05E-15	&	9E-16	&       5.38E-15        &       9E-16       & 74c     &       6.27E-15        &       8E-16       &       3.99E-15        &       7E-16       \\ 
33b	&	3.56E-15	&	9E-16	&       2.60E-15        &       8E-16       & 74d     &       6.6E-15         &       7E-16       &       4.65E-15        &       7E-16      \\ 
33c	&	4.65E-14	&	3E-15	&       4.46E-14        &       3E-15       & 75a     &       3.75E-14        &       1E-15       &       1.86E-14        &       9E-16       \\ 
33d	&	1.45E-15	&	8E-16	&       6.50E-16        &       8E-16       & 75b     &       7.87E-15        &       6E-16       &       6.96E-15        &       6E-16       \\ 
34c	&	5.54E-14	&	1E-14	&       4.47E-14        &       1E-14       & 75c     &       1.24E-14        &       7E-16       &       8.22E-15        &       6E-16       \\ 
35a	&	1.57E-14	&	2E-15	&       1.34E-14        &       2E-15       & 75d     &       3.27E-14        &       1E-15       &       3.12E-14        &       1E-15       \\ 
35b	&	8.31E-15	&	2E-15	&       8.75E-15        &       2E-15       & 75f     &       2.42E-15        &       4E-16       &       \               &       \       \\ 
35c	&	9.20E-15	&	2E-15	&       9.01E-15        &       2E-15       & 76a     &       1.28E-14        &       2E-15       &       1.25E-14        &       2E-15       \\ 
35d	&	1.03E-14	&	2E-15	&       9.70E-15        &       2E-15       & 76b     &       1.09E-14        &       2E-15       &       \               &       \       \\ 
35e	&	1.93E-15	&	2E-15	&       2.58E-15        &       2E-15       & 76c     &       1.01E-14        &       2E-15       &       \               &       \       \\ 
35f	&	1.73E-15	&	2E-15	&       1.20E-15        &       2E-15       & 76d     &       1.14E-14        &       2E-15       &       \               &       \       \\ 
37a	&	5.45E-14	&	9E-15	&       5.07E-14        &       9E-15       & 76f     &       7.08E-15        &       2E-15       &       5.02E-15        &       2E-15       \\ 
37b	&	3.01E-14	&	9E-15	&       2.74E-14        &       9E-15       & 82b     &       5.52E-15        &       1E-15       &       8.79E-16        &       1E-15       \\ 
37c	&	6.25E-15	&	9E-15	&       5.36E-15        &       9E-15       & 82c     &       3.98E-14        &       2E-15       &       3.47E-14        &       2E-15       \\ 
37d	&	1.98E-14	&	1E-14	&       1.84E-14        &       1E-14       & 83b     &       1.46E-15        &       1E-15       &       2.32E-15        &       1E-15       \\ 
37e	&	4.31E-15	&	1E-14	&       3.64E-15        &       1E-14       & 83c     &       3.46E-14        &	      3E-15       &       	        	  3.26E-14	&       3E-15       \\ 
38a	&	4.34E-14	&	5E-15	&       3.89E-14        &       5E-15       & 92c     &       5.66E-14        &       3E-15       &     5.28E-14	&    3E-15          \\  
38b	&	4.87E-14	&	6E-15	&       4.55E-14        &       6E-15       \\
38c	&	2.24E-14	&	4E-15	&       2.13E-14        &       4E-15       \\ 
43a	&	6.15E-14	&	4E-15	&       5.34E-14        &       4E-15       \\ 
43b	&	5.42E-14	&	4E-15	&       4.93E-14        &       3E-15       \\ 
43c	&	3.07E-14	&	2E-15	&       2.63E-14        &       2E-15       \\ 
45a	&	4.64E-14	&	2E-15	&       4.95E-14        &       2E-15       \\ 
45b	&	8.54E-15	&	2E-15	&       7.59E-15        &       2E-15       \\ 
45c	&	3.9E-15		&	2E-15	&       5.03E-15        &       2E-15       \\ 
46a	&	5.48E-15	&	7E-16	&       4.76E-15        &       7E-16       \\ 
46b	&	3.47E-15	&	7E-16	&       2.95E-15        &       6E-16       \\ 
46c	&	1.63E-15	&	6E-16	&       1.36E-15        &       6E-16       \\ 
46d	&	3.55E-16	&	6E-16	&       7.21E-17        &       6E-16       \\ 
49a	&	5.47E-14	&	2E-15	&       5.21E-14        &       2E-15       \\ 
49b	&	1.22E-13	&	4E-15	&       1.20E-13        &       4E-15       \\ 
49c	&	2.66E-14	&	2E-15	&       2.47E-14        &       2E-15       \\ 
49d	&	1.25E-14	&	1E-15	&       1.28E-14        &       1E-15       \\ 
53a	&	2.1E-13		&	4E-15	&       1.84E-13        &       4E-15       \\ 
53b	&	1.69E-14	&	4E-15	&       1.57E-14        &       4E-15       \\ 
53c	&	1.05E-13	&	4E-15	&       9.76E-14        &       4E-15       \\ 
56a	&	3.15E-14	&	1E-15	&       2.65E-14        &       1E-15       \\ 
56b	&	1.59E-13	&	3E-15	&       1.58E-13        &       3E-15       \\ 
56d	&	2.98E-14	&	1E-15	&       2.81E-14        &       1E-15       \\ 
56e	&	8.46E-15	&	9E-16	&       8.64E-15        &       9E-16       \\ 
66b	&	5.87E-14	&	4E-15	&       7.98E-15        &       5E-16       \\ 
69a	&	1.06E-14	&	2E-15	&       1.18E-14        &       2E-15       \\ 
69b	&	4.62E-14	&	4E-15	&       4.46E-14        &       4E-15       \\ 
70e	&	2.90E-14	&	1E-15	&       2.43E-14        &       1E-15       \\ 
70g	&	1.3E-14		&	7E-16	&       1.03E-15        &       6E-16       \\ 
71a	&	1.64E-13	&	8E-15	&       1.49E-13        &       7E-15       \\ 
71b	&	6.96E-15	&	7E-16	&       1.12E-14        &       8E-16       \\ 
71c	&	3.22E-14	&	2E-15	&       2.96E-14        &       1E-15       \\ 
72a	&	8.44E-15	&	2E-15	&       8.40E-15        &       2E-15       \\ 
72b	&	3.95E-15	&	2E-15	&       3.94E-15        &       2E-15       \\ 
72c	&	2.92E-15	&	2E-15	&       2.93E-15        &       2E-15       \\ 
72d	&	4.14E-15	&	2E-15	&       4.27E-15        &       2E-15       \\  
\noalign{\smallskip}
\hline
\hline
\end{tabular} 
\end{flushleft}
\end{table*}
\newpage
\normalsize
\begin{table*}[h]
{\bf Table 9:} Upper Limit to Flux and Luminosity
\begin{flushleft}
\begin{tabular}{ccccc}
\hline
\hline
\noalign{\smallskip}
\noalign{\smallskip}
Galaxy & $f_{3\sigma}$ & $\sigma$ & $L_{3\sigma}$ & $\sigma$ \\
\noalign{\smallskip} 
\noalign{\smallskip}
 \     & $\left(erg~cm^{-2}s^{-1} \right)$ & $\left(erg~cm^{-2}s^{-1} \right)$ & $\left(erg~s^{-1} \right)$ & $\left(erg~s^{-1} \right)$ \\
\noalign{\smallskip}
\noalign{\smallskip}
\hline
\noalign{\smallskip}
34a	&	8.08E-17	&	1E-14	&	7.51E+37	&	9E+39	\\ 
34d	&	7.19E-17	&	8E-15	&	6.41E+37	&	8E+39	\\ 
56c	&	1.29E-16	&	9E-16	&	9.76E+37	&	7E+38	\\ 
66a	&	7.82E-17	&	5E-18	&	3.91E+38	&	2E+37	\\ 
66c	&	7.82E-17	&	5E-18	&	3.96E+38	&	2E+37	\\ 
66d	&	7.82E-17	&	5E-18	&	3.98E+38	&	2E+37	\\ 
68a	&	1.45E-16	&	2E-15	&	7.69E+36	&	8E+37	\\ 
68b	&	1.50E-16	&	2E-15	&	1.18E+37	&	1E+38	\\ 
69c	&	1.86E-16	&	2E-15	&	1.55E+38	&	2E+39	\\ 
69d	&	2.50E-16	&	3E-15	&	2.40E+38	&	3E+39	\\ 
70f	&	2.56E-16	&	6E-16	&	1.10E+39	&	3E+39	\\ 
74e	&	1.19E-16	&	6E-16	&	1.81E+38	&	8E+38	\\ 
75e	&	1.30E-16	&	4E-16	&	2.27E+38	&	7E+38   \\ 
76e	&	2.39E-16	&	2E-15	&	2.94E+38	&	2E+39	\\ 
82a	&	1.72E-16	&	1E-15	&	2.47E+38	&	2E+39	\\ 
82d	&	1.74E-16	&	1E-15	&	2.74E+38	&	2E+39	\\
83a     &       2.20E-16        &	2E-15	&	6.18E+38        &	3E+39   \\
83d	&	2.06E-16	&	1E-15	&	5.73E+38	&	3E+39	\\ 
83e	&	2.07E-16	&	1E-15	&	5.81E+38	&	3E+39	\\ 
92b	&	2.32E-16	&	3E-15	&	8.84E+37	&	1E+39	\\ 
92d	&	1.73E-16	&	2E-15	&	8.69E+37	&	1E+39	\\ 
92e	&	1.73E-16	&	2E-15	&	8.62E+37	&	1E+39	\\ 
\noalign{\smallskip}
\hline
\hline
\end{tabular} 
\end{flushleft}
\end{table*}

\section {Star formation Rate}

So far this is the largest $H_\alpha$ catalogue of HCG galaxies  having $H_\alpha$ calibrated fluxes. Such a sample constitutes a powerful tool to perform quantitative analysis on the recent star formation rate inside HCG galaxies.
We have derived the {\em SFR} for the galaxies of our sample using the results of Kennicutt (1983), which relate the SFR to $H_{\alpha}$ luminosity through the relation:
\begin{equation} 
SFR(total)={{L(H_\alpha)}\over{1.12\cdot10^{41} {\rm erg~~s^{-1}}}} M_\odot yr^{-1} 
\end{equation} 
where a Salpeter initial mass function with an upper mass cutoff of 100 
$M_\odot$ has been assumed. {\em SFR} inferred from luminosities for the 73 galaxies detected and for 22 upper limits estimated is shown in Table 10 as follows:\\
Col. 1: Name of detected galaxies;\\
Col. 2: {\em SFR} inferred from isophotal luminosities $L_{iso}$ corrected for Galactic Extinction, $SFR_{iso}$ (2) ;\\
Col. 3: {\em SFR} inferred from isophotal luminosities, corrected  for both Galactic and Internal Extinction, $SFR_{iso}$ (3);\\
Col. 5: Name of galaxies for which we have computed the upper limits;\\
Col. 6: {\em SFR} inferred from  upper limit to luminosity, corrected for Galactic Extinction, $SFR_{ul}$ (2);\\
Col. 7: {\em SFR} inferred from  upper limit to luminosity, corrected  for both Galactic and Internal Extinction, $SFR_{ul}$ (3);\\
\begin{table*}[h]
{\bf Table 10:} SFR of galaxies in the sample
\small
\begin{flushleft}
\begin{tabular}{ccccccccc}
\hline
\hline
\noalign{\smallskip}
\noalign{\smallskip}
Galaxy & $SFR_{iso}$ (2) & $SFR_{iso}$ (3)  & Galaxy & $SFR_{iso}$ (2) & $SFR_{iso}$ (3)  & Upper Limits & $SFR_{ul}$ (2) & $SFR_{ul}$ (3)\\
\noalign{\smallskip} 
\noalign{\smallskip}
 \     & $M_\odot$ yr$^{-1}$ & $M_\odot$ yr$^{-1}$  &  \     & $M_\odot$ yr$^{-1}$ & $M_\odot$ yr$^{-1}$  &   \    &  $M_\odot$ yr$^{-1}$ & $M_\odot$ yr$^{-1}$\\
\noalign{\smallskip}
\noalign{\smallskip}
\hline
\noalign{\smallskip}
2a	&	1.4465	&	1.7363	&  53c	&	0.3780	&	0.4860	&	34a	&	0.0014	&	0.0014	\\
2b	&	1.1584	&	1.1584	&  56a	&	0.1572	&	0.1842	&	34d	&	0.0012	&	0.0011	\\
33a	&	0.0892	&	0.0892	&  56b	&	1.0621	&	1.0621	&	56c	&	0.0009	&	0.0009	\\
33b	&	0.0556	&	0.0556	&  56d	&	0.2085	&	0.2085  &	66a	&	0.0037	&	0.0037	\\
33c	&	0.6754	&	0.7032	&  56e	&	0.0530	&	0.0530	&	66c	&	0.0038	&	0.0038	\\
33d	&	0.0196	&	0.0196	&  66b	&	2.8838	&	2.8838	&	66d	&	0.0038	&	0.0038	\\
34c	&	0.6843	&	0.8435	&  69a	&	0.0697	&	0.0844	&	68a	&	7E-05	&	7E-05	\\
35a	&	0.4360	&	0.4360	&  69b	&	0.3669	&	0.5241	&	68b	&	0.0001	&	0.0001	\\
35b	&	0.2569	&	0.2569	&  70e	&	0.8952	&	1.4116	&	69c	&	0.0015	&	0.0015	\\
35c	&	0.2893	&	0.2893	&  70g	&	0.4604	&	0.7295	&	69d	&	0.0023	&	0.0023	\\
35d	&	0.2633	&	0.2845	&  71a	&	1.3874	&	2.0588	&	70f	&	0.0105	&	0.0154	\\
35e	&	0.0695	&	0.0695	&  71b	&	0.0836	&	0.0926	&	74e	&	0.0019	&	0.0019	\\
35f	&	0.0375	&	0.0375	&  71c	&	0.2055	&	0.2877	&	75e	&	0.0024	&	0.0033	\\
37a	&	0.2623	&	0.2623	&  72a	&	0.1457	&	0.1909	&	76e	&	0.0031	&	0.0031	\\
37b	&	0.1273	&	0.1748	&  72b	&	0.0668	&	0.0668	&	82a	&	0.0024	&	0.0024	\\
37c	&	0.0316	&	0.0316	&  72c	&	0.0584	&	0.0584	&	82d	&	0.0027	&	0.0027	\\
37d	&	0.0757	&	0.0757	&  72d	&	0.0693	&	0.0693	&	83a	&	0.0072	&	0.0072	\\
37e	&	0.0153	&	0.0153	&  74a	&	0.5054	&	0.5054	&	83d	&	0.0067	&	0.01	\\
38a	&	0.3385	&	0.3614	&  74b	&	0.1185	&	0.11845	&	83e	&	0.0068	&	0.0068	\\
38b	&	0.3933	&	0.5111	&  74c	&	0.0959	&	0.0959  &	92b	&	0.0011	&	0.0013	\\
38c	&	0.1901	&	0.1901	&  74d	&	0.0962	&	0.0962  &	92d	&	0.0011	&	0.0011	\\
43a	&	0.6562	&	0.6701	&  75a	&	0.5775	&	0.5775  &	92e	&	0.0011	&	0.0017	\\
43b	&	0.5796	&	0.7323	&  75b	&	0.1153	&	0.1331  &	\	&	\	&	\	\\
43c	&	0.2765	&	0.2765	&  75c	&	0.1698	&	0.1698  &	\	&	\	&	\	\\
45a	&	1.9142	&	2.3940	&  75d	&	0.5578	&	0.7059  &	\	&	\	&	\	\\
45b	&	0.381	&	0.381	&  75f	&	0.0289	&	0.0289  &	\	&	\	&	\	\\
45c	&	0.1437	&	0.1747	&  76a	&	0.1475	&	0.1752  &	\	&	\	&	\	\\
46a	&	0.0373	&	0.0373	&  76b	&	0.1432	&	0.1432  &	\	&	\	&	\	\\
46b	&	0.0267	&	0.0267	&  76c	&	0.1524	&	0.1524  &	\	&	\	&	\	\\
46c	&	0.0106	&	0.0106	&  76d	&	0.1483	&	0.1483  &	\	&	\	&	\	\\
46d	&	0.0021	&	0.0021	&  76f	&	0.0740	&	0.1004  &	\	&	\	&	\	\\
49a	&	0.5545	&	0.8761	&  82b	&	0.0505	&	0.0728  &	\	&	\	&	\	\\
49b	&	1.2973	&	1.7712	&  82c	&	0.383	&	0.383   &	\	&	\	&	\	\\
49c	&	0.2576	&	0.2576	&  83b	&	0.048	&	0.048   &	\	&	\	&	\	\\
49d	&	0.1352	&	0.1352	&  83c	&	1.1171	&	1.3999  &	\	&	\	&	\	\\
53a	&	0.6193	&	0.6194	&  92c	&	0.3387	&	0.5007  &	\	&	\	&	\	\\
53b	&	0.06160	&	0.0616	&   \   &       \       &       \       &	\	&	\	&	\	\\
\noalign{\smallskip} 
\hline
\hline
\end{tabular} 
\end{flushleft}
\end{table*}
\normalsize
In Figure 5 we show the distribution of $SFR_{iso}$ computed taking into account (dotted line) and without taking into account (solid line) Internal Extinction. The two distributions are quite similar, as confirmed also by a Kolmogorov-Smirnov test.
\begin{figure}[h]
\centerline{\psfig{figure=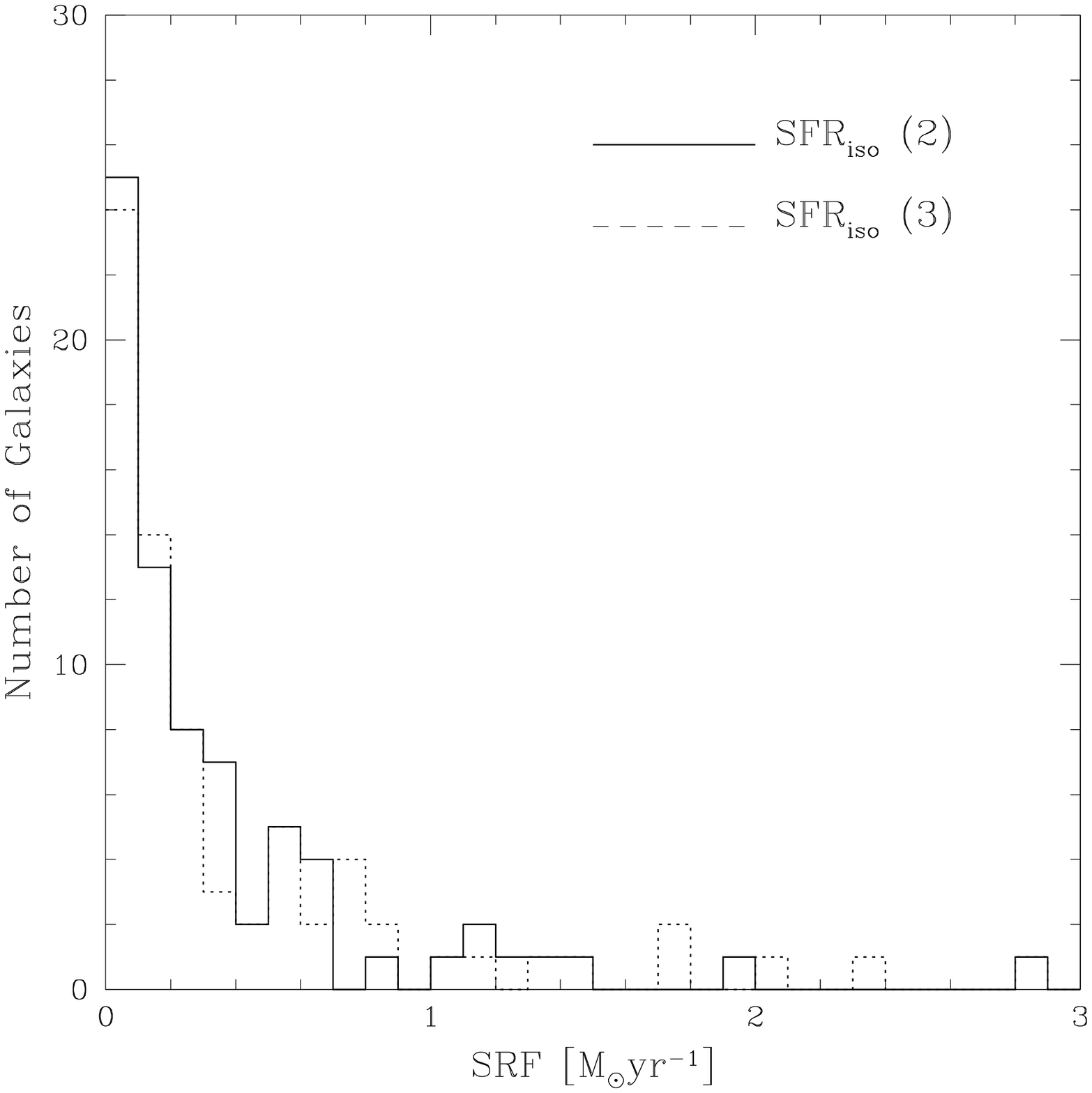,height=80mm,bbllx=70mm,bblly=60mm,bburx=140mm,bbury=250mm}}
\bigskip
\bigskip
{{\bf Figure 5:} Distribution of SFR for the 73 detected galaxies derived from $L_{iso}$. The solid histogram represents the $SFR_{iso}$ (2) and the dotted histogram represents the  $SFR_{iso}$ (3) (see Table 10). The width of each bin is 0.1 [$M_\odot$ yr$^{-1}$].}
\end{figure}
\vskip 0.5truecm

\section {Discussion}

Figures 6 to 11 show the continuum and $H_\alpha$ maps of some of the HCG galaxies of our sample. 
A isocontour plot is also shown for the largest galaxies of these groups.
In the  $H_\alpha$ images we have removed  residuals of  field stars for clarity and we display the galaxy flux  one sigma above the background. 
The scale of the axes are in pixel units and the field of view is 5.12x5.12 arcmin.
East is on the top and North on the left.
For each image the name of HCG is shown in the caption, while the name of the galaxies are reported 
in the figure.
In the following we present a brief description of the groups and galaxies in figures 6 to 11.
The values regarding the flux of the lowest isocontours are corrected for Galactic Extinction.
\vskip 0.5truecm
{\bf HCG2-} This group consists of a triplet of galaxies with accordant redshifts (galaxies a, and c) plus a fainter member (galaxy d) which has a  higher redshift. In Figure 6  are included only the galaxies a and b, for which we have estimated the $H_\alpha$ fluxes. Galaxy a (late type barred spiral) is  brighter in $H_\alpha$ than b (compact irregular), which is also a infrared source. In the $H_\alpha$ map some knots are resolved in the disk of galaxy a. For galaxy b we detect a strong $H_\alpha$ emission in the center and no emission in the outer part, as previously noted by Vilchez \& Iglesias Paramo \cite{Vil:Para}.
The estimated $SFR_{iso}$ for a and b are respectively 1.45 and 1.16 $M_\odot$ yr$^{-1}$.
In the lowest panel of Figure 6 the isocontour plot, showing the shape and 
the orientation of the $H_\alpha$ emission, is given.
The lowest contour is at 1$\sigma$ above the background, 
corresponding to an $H_\alpha$ emission of  $6.27 \cdot 10^{-17}~erg~cm^2s^{-1}arcsec^{-2}$.
The interval among the contours is 3$\sigma$.
The $H_\alpha$ emitting areas have an extension of about  
1643.6 and 315.8 $arcsec^2$  for a and b galaxies respectively.
\vskip 0.5truecm
{\bf HCG37-} This is a compact group with five accordant galaxies: a and b are the dominant galaxies of the group. They are radio sources, as galaxy d. The group has a high velocity dispersion (398.1 $km~s^{-1}$) and mass-to-light ratio (123 $M_\odot\L_\odot$), and a short crossing time (0.0054 $Ht_c$). The  $H_\alpha$ brightest source of the group is galaxy a. This is a blue elliptical galaxy with a rapidly rotating central disk of ionized gas (Rubin et al. \cite{Rub}). Galaxy b is an edge-on spiral with an intensive $H_\alpha$ emission in the center. This galaxy is  also an infrared source. 
Galaxies c (SOa), d (SBdm) and e (E0) are all  fainter $H_\alpha$ emitters than galaxies a and b.
The $SFR_{iso}$ estimated for a, b, c, d and e galaxies are respectively 0.26, 0.13, 0.03, 0.08 and 
0.02 $M_\odot$ yr$^{-1}$.
The lowest panel of Figure 7 presents the isocontours for the largest 
 galaxies of the group:
the extensions of $H_\alpha$ emission are of about 235.2, 217.1, 46.7, 98.4 and 28.4 $arcsec^2$ 
 for a, b,c, d and e galaxies respectively.
The lowest contour is at 1$\sigma$ above the background, 
that is $H_\alpha$ emission higher than $3.64 \cdot 10^{-17}~erg~cm^2s^{-1}arcsec^{-2}$,
while the interval among the contours is 3$\sigma$.
\vskip 0.5truecm
{\bf HCG38-} This group contains the interacting pair Arp 237 (galaxies b and c) with one other galaxy at a similar redshift (galaxy a), plus a fainter high-redshift galaxy (d). Galaxy a is a spiral showing an $H_\alpha$ emission  more intense in the center than in the outer disk. The $H_\alpha$ brightest galaxy b (late type barred spiral) is in the interacting pair and it is an infrared source.  In galaxy b we reveal a strong $H_\alpha$ emission in the central part of the galaxy and some resolved knots throughout its arm placed in the direction opposite to galaxy c. This last galaxy is of irregular type and it is the $H_\alpha$ dimmest galaxy of the group.
The estimated $SFR_{iso}$ for a, b and c are respectively 0.34, 0.39 and 0.19 $M_\odot$ yr$^{-1}$.
The isocontour plots in Figure 8 show the $H_\alpha$ emission higher than 
$7.22 \cdot 10^{-17}~erg~cm^2s^{-1}arcsec^{-2}$ (1$\sigma$ above the background). 
The interval among the contours is 1$\sigma$.
The extensions of $H_\alpha$ emission are of about 156.3, 150.4 
and 60.0 $arcsec^2$  for a, b and c  galaxies respectively.
\vskip 0.5truecm
{\bf HCG46-} This group consists of four early-type galaxies.
The velocity dispersion and mass-to-light ratio of the group is relatively high (respectively 323.6 $km~s^{-1}$ and 478.6 $M_\odot/L_\odot$).
Galaxies b and c appear to be in contact in the continuum image, but not in the $H_\alpha$ map.  Two features are in common to all the galaxies of the group: they show a faint $H_\alpha$ emissions that seems confined to the bulge of galaxies.
The estimated $SFR_{iso}$ for a, b, c and d are respectively 0.04, 0.03, 0.01 and 0.002 $M_\odot$ yr$^{-1}$.
The  extensions of H$_\alpha$ emission at 1$\sigma$ above the background are reported in Table 7.
\vskip 0.5truecm
{\bf HCG49-} This is a small and very compact group with four accordant galaxies. Its median galaxy separation is only of 12.3 $h^{-1} kpc$ and its velocity dispersion is so low (lower than the uncertainties in the velocity measurements) that no estimate can be made of its mass-to-light ratio (Hickson \cite{Hik}). Galaxies a and b are spiral, while c is an irregular and d an elliptical. Galaxy b is the $H_\alpha$ brightest source of the group, while galaxies a, c and d have comparable $H_\alpha$ 
emission among them.
The estimated $SFR_{iso}$ for a, b, c and d are respectively 0.56, 1.3, 0.26 and 0.14 
$M_\odot$ yr$^{-1}$.
In the lowest panel of Figure 10 the isocontour plot is shown.
The lowest contour is at 1$\sigma$ above the background, 
corresponding to an $H_\alpha$ emission of $9.44 \cdot 10^{-17}~erg~cm^2s^{-1}arcsec^{-2}$.
The interval among the contours is 2$\sigma$.
The extensions of H$_\alpha$ emission are of about 72.8, 96.3, 48.9 and 27 $arcsec^2$  for a, b, c 
and d galaxies respectively.
\vskip 0.5truecm
{\bf HCG74-} This group contains five early-type accordant galaxies: a, b, d are elliptical and c and e are lenticular. Galaxy a is the dominant one with two very close companions (b and c) and it is also a radio source. All galaxies show an $H_\alpha$ emission confined to their center. We have not revealed e galaxy for which we have estimated the upper limit.
The estimated $SFR_{iso}$ for a, b, c and d detected galaxies are respectively 0.51, 0.12, 0.1 and 
0.1 $M_\odot$ yr$^{-1}$.
In the lowest panel of Figure 11 the isocontour plots for a, b and c galaxies 
are shown.
The lowest contour is at 1$\sigma$ above the background ($7.52 \cdot 10^{-17}~erg~cm^2s^{-1}arcsec^
{-2}$), while the interval among the contours is 1$\sigma$.
The extensions of H$_\alpha$ emission are 61.38, 40.59, 21.96 and 23.04 $arcsec^2$  for a, b, c and d galaxies respectively.
\vskip 1truecm

\section {Summary}

We have obtained $H_\alpha$ fluxes and luminosities for a sample of 95 galaxies from calibrated observations of 31 HCGs. The sample thus collected comprises 75$\%$ of the accordant galaxies of the observed groups and it represents the largest $H_\alpha$ selected sample of HCG galaxies so far having calibrated fluxes. By the estimated $L_{H_\alpha}$ we have obtained the star formation rate of the sample galaxies.
In a following paper (Severgnini \& Saracco, 1999) we will combine the results obtained from the data presented here 
with dynamical, morphological and broad band photometrical data from the literature (Hickson \cite{Hic}, Hickson \cite{Hik}, Rood \& Struble \cite{Rod:Str}) to show that the $H_\alpha$ luminosity of galaxies and hence their current star formation rate are affected by the dynamics of groups in which they reside.\\
Further analysis based on these $H_\alpha$ data will allow us to study the rate of interaction and merger phenomena occurring in HCGs, yielding new insights about the formation and evolution of these systems.
\newpage
\begin{figure*}[h]
\centerline{\psfig{figure=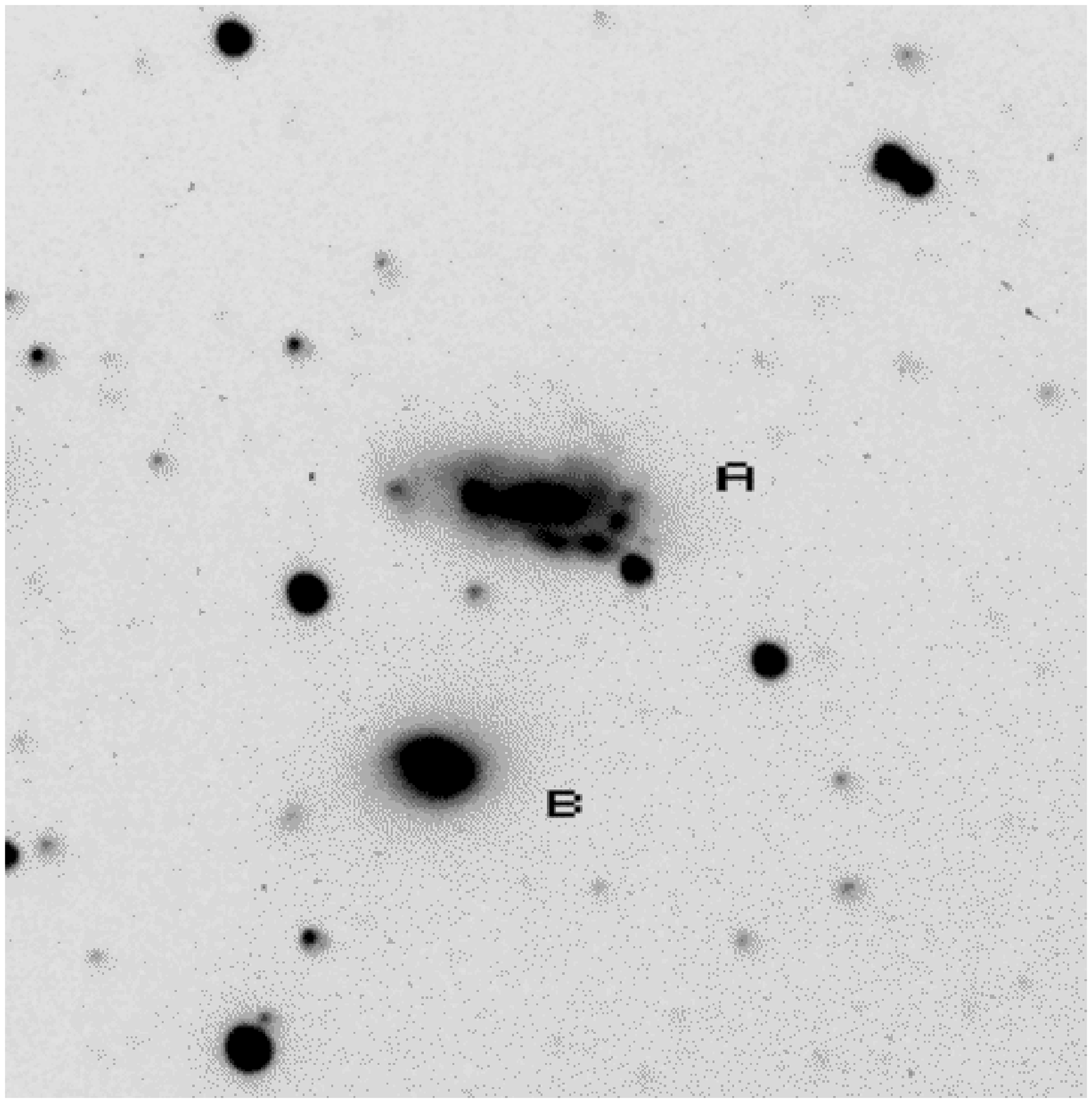,height=75mm,bbllx=70mm,bblly=60mm,bburx=140mm,bbury=250mm}}
\centerline{\psfig{figure=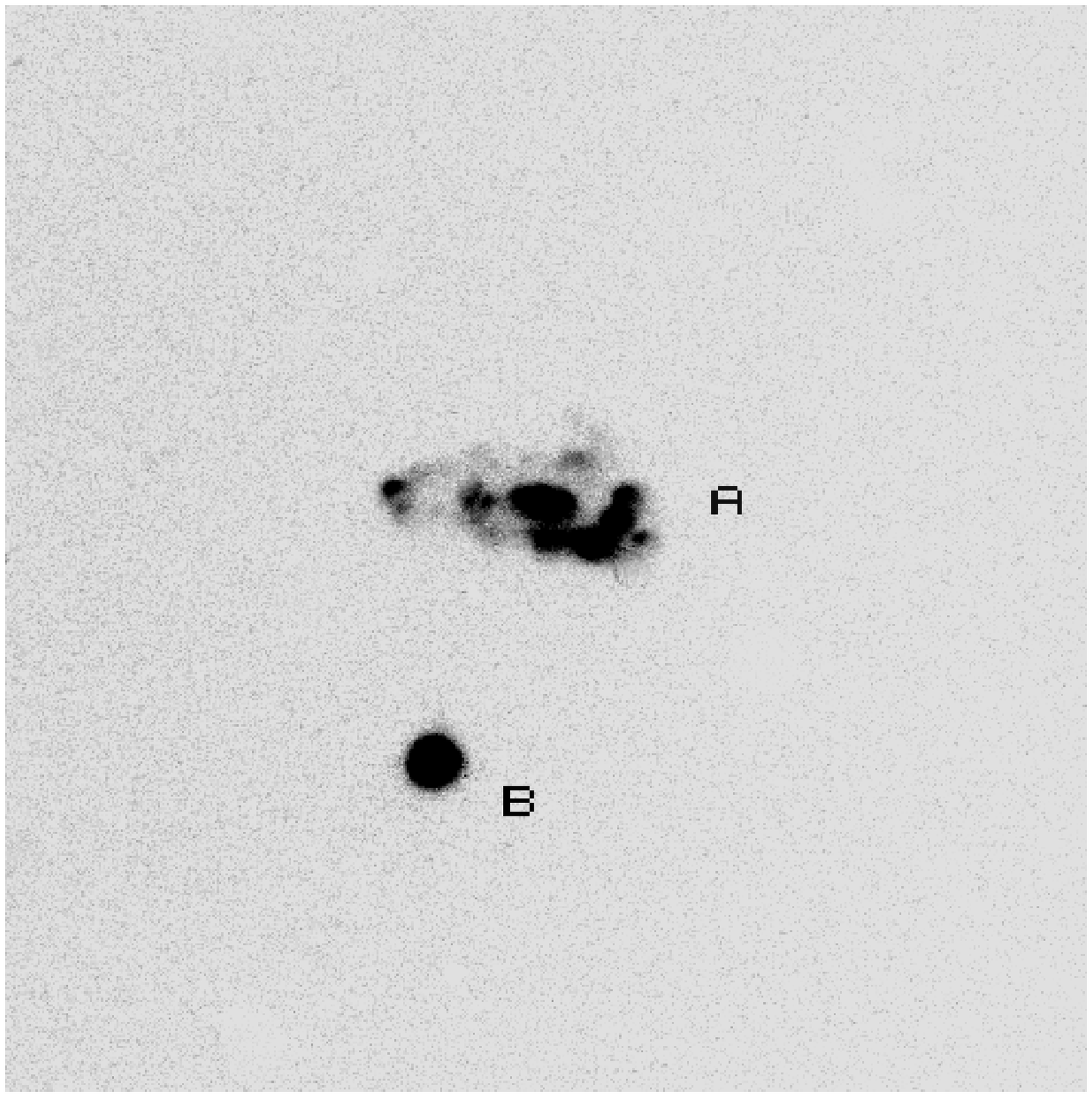,height=75mm,bbllx=70mm,bblly=60mm,bburx=140mm,bbury=250mm}}
\vskip - 5truecm
\hskip -0.25truecm
\centerline{\psfig{figure=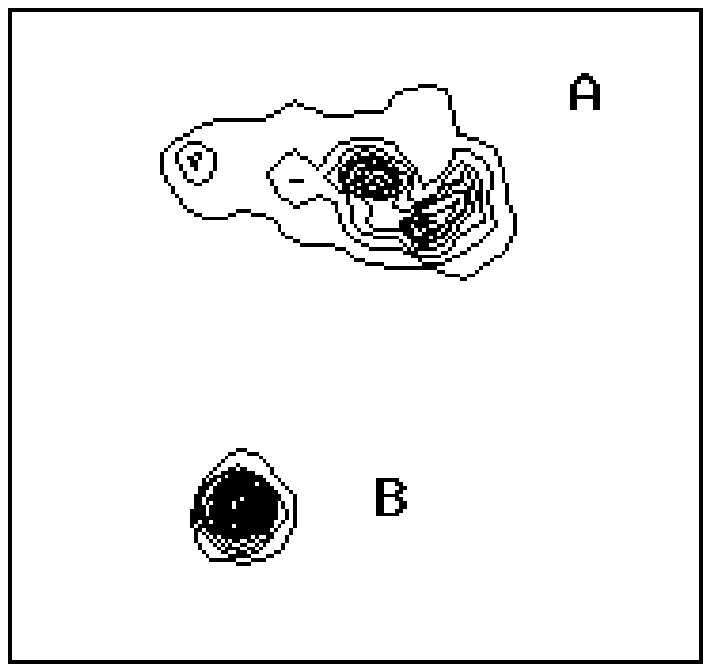,height=170mm,bbllx=70mm,bblly=60mm,bburx=140mm,bbury=250mm}}
\vskip -4truecm
{\bf Figure 6:} Continuum (up), $H_\alpha$ map (middle) and zoomed isocontour 
map (down) of galaxies A and B of  HCG2.
The lowest contour is at 1$\sigma$ ($6.27 \cdot 10^{-17} erg~cm^2s^{-1}arcsec^{-2}$) above the background.
The interval among the contours is 3$\sigma$. 
\end{figure*}
\newpage
\begin{figure*}[h]
\centerline{\psfig{figure=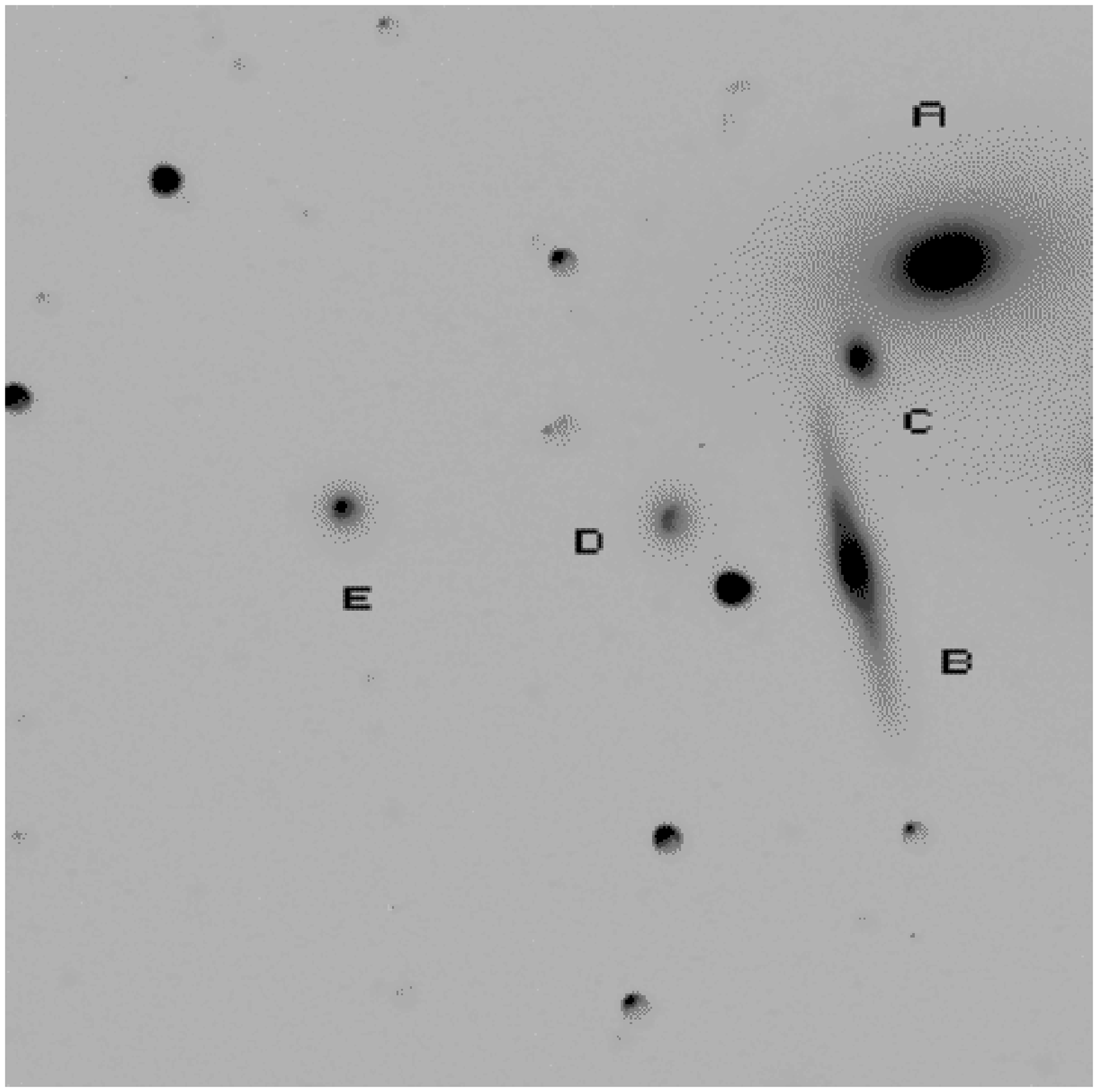,height=75mm,bbllx=70mm,bblly=60mm,bburx=140mm,bbury=250mm}}
\centerline{\psfig{figure=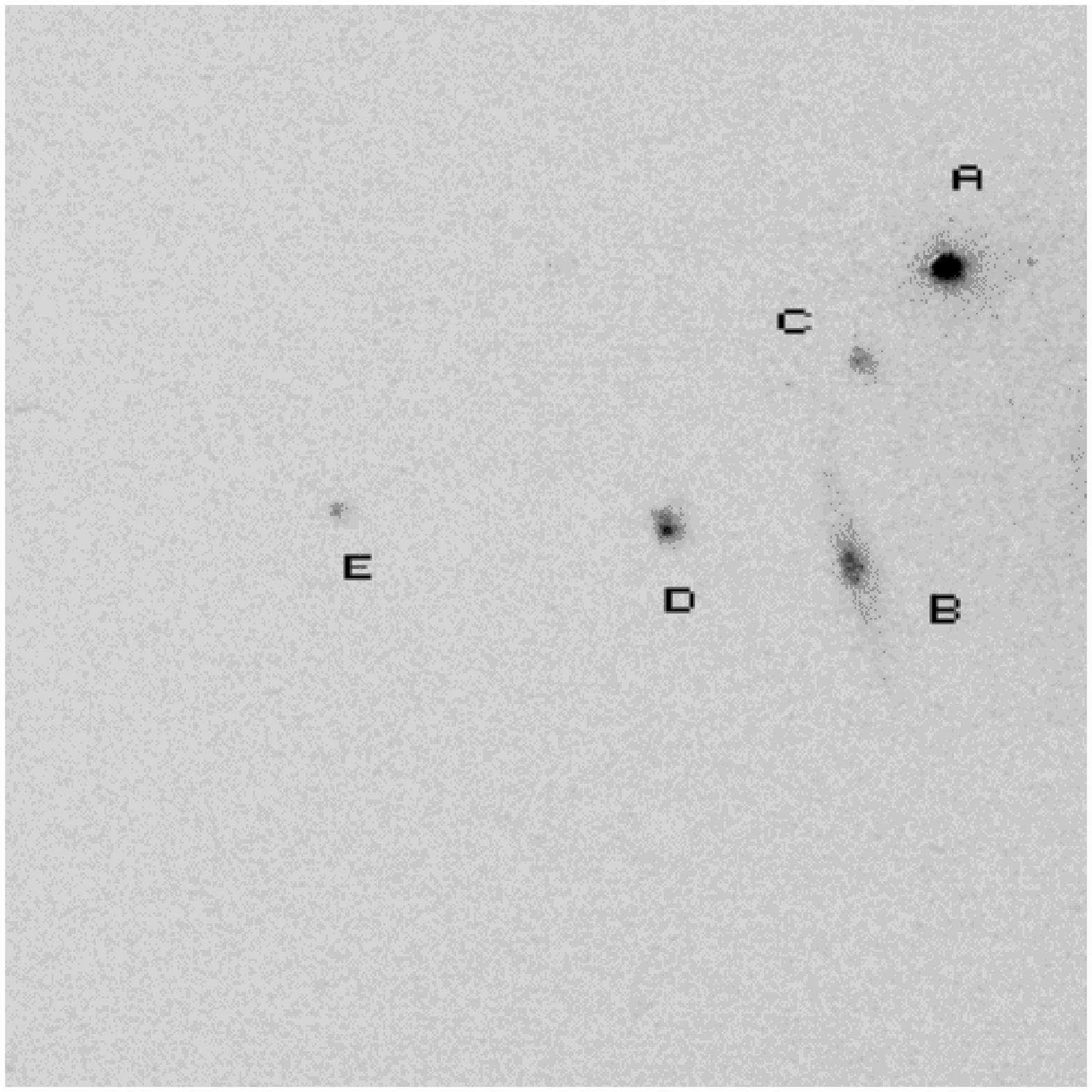,height=75mm,bbllx=70mm,bblly=60mm,bburx=140mm,bbury=250mm}}
\vskip -5.5truecm
\hskip -0.25truecm
\centerline{\psfig{figure=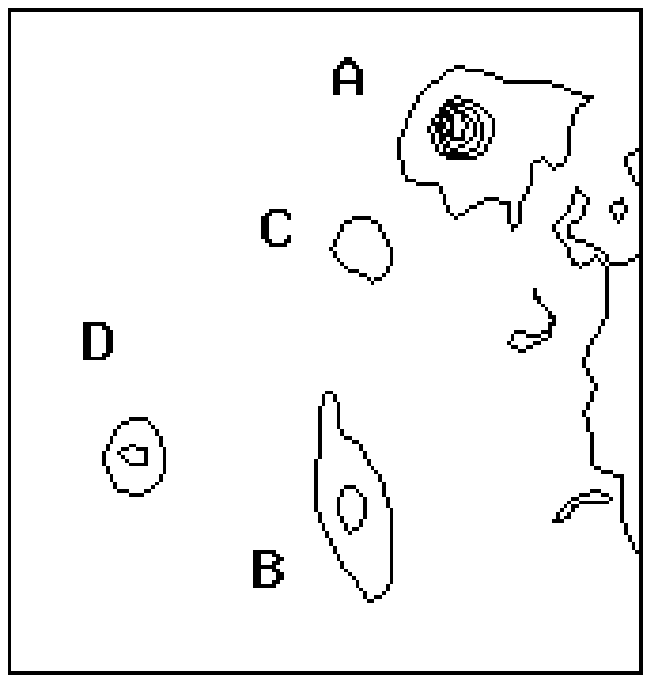,height=180mm,bbllx=70mm,bblly=60mm,bburx=140mm,bbury=250mm}}
\vskip -4.5truecm
{\bf Figure 7:} Continuum (up), $H_\alpha$ map (middle) and zoomed isocontour 
map (down)  of HCG37.
The isocontour plots are given for the largest galaxies only (see Table 7).
The lowest contour is at 1$\sigma$ ($3.64 \cdot 10^{-17}~erg~cm^2s^{-1}arcsec^{-2}$) above the background.
The interval among the contours are 3$\sigma$. 
\end{figure*}
\newpage
\begin{figure*}[h]
\centerline{\psfig{figure=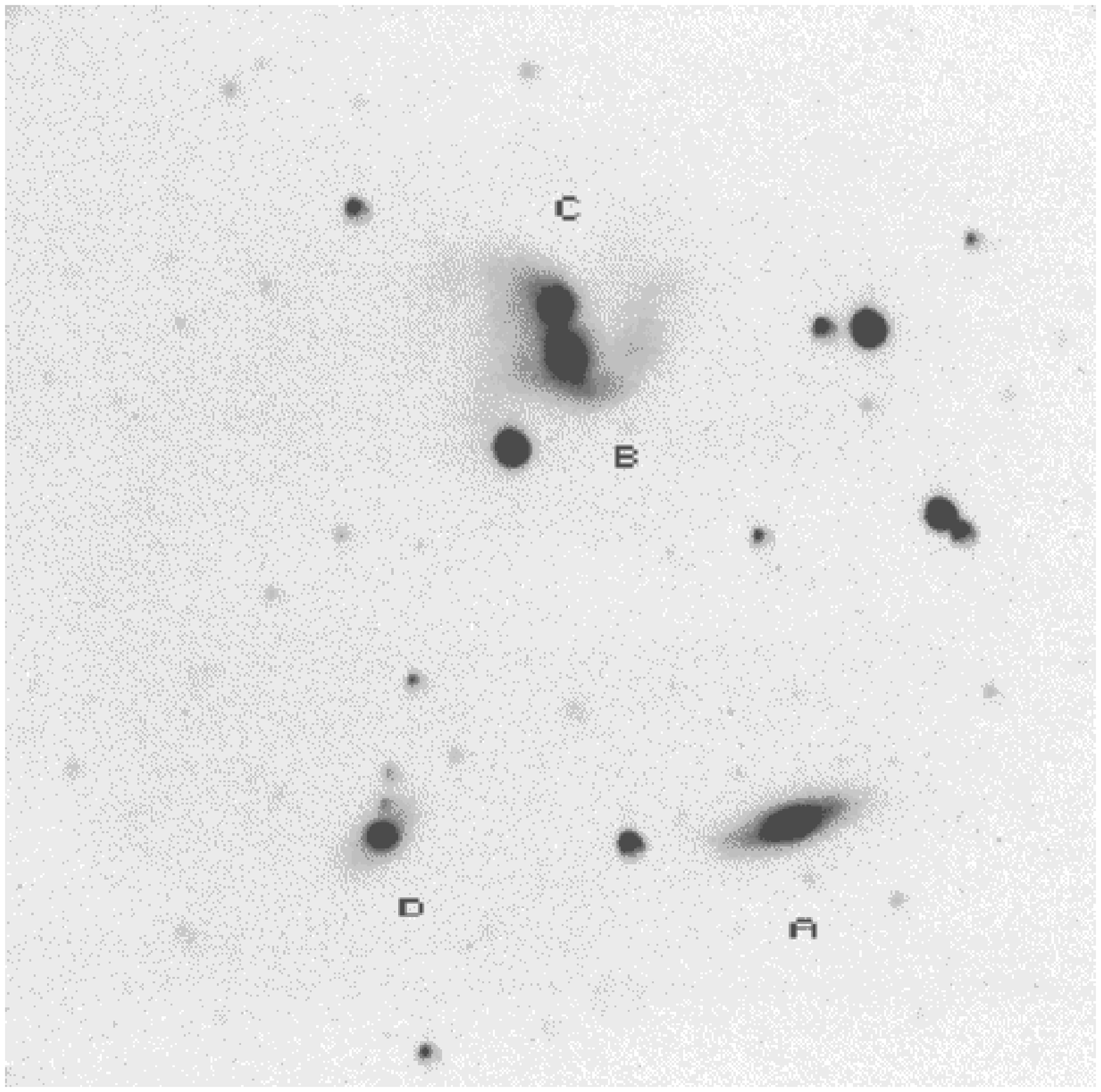,height=75mm,bbllx=70mm,bblly=60mm,bburx=140mm,bbury=250mm}}
\centerline{\psfig{figure=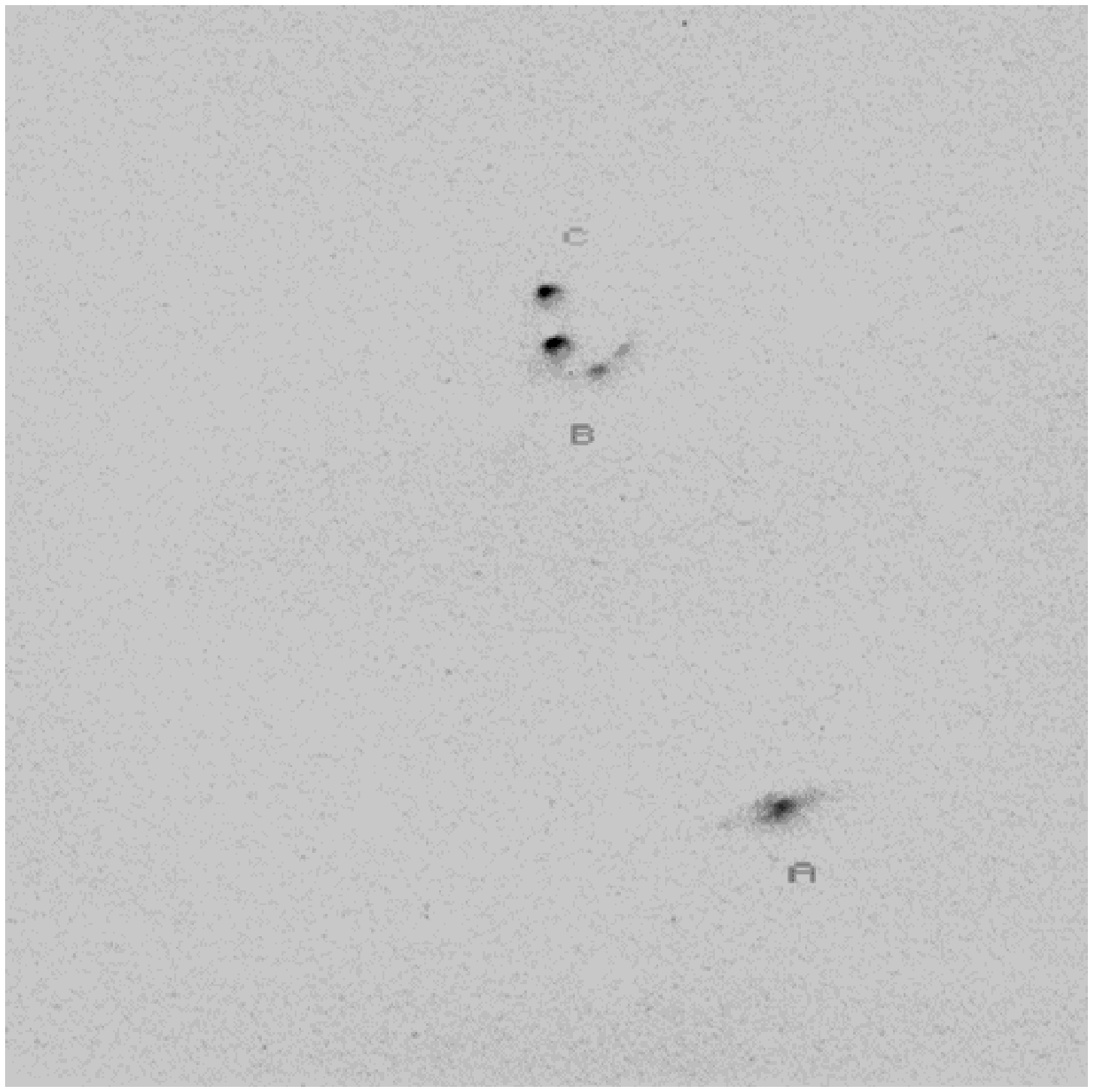,height=75mm,bbllx=70mm,bblly=60mm,bburx=140mm,bbury=250mm}}
\vskip -7truecm
\hskip -0.5truecm
\psfig{figure=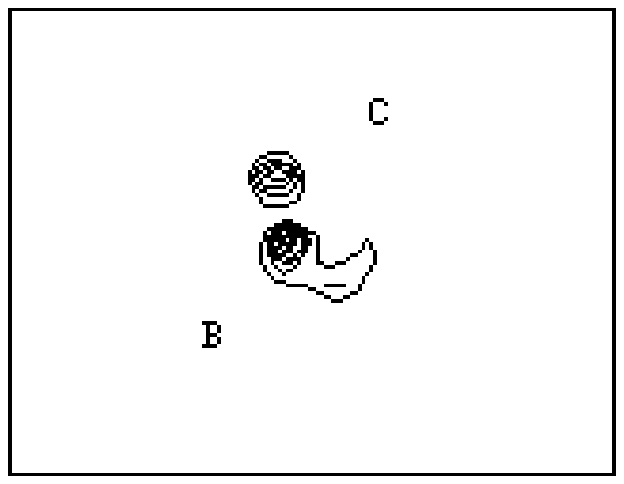,height=200mm,bbllx=70mm,bblly=60mm,bburx=140mm,bbury=250mm}
\vskip -22.2truecm
\hskip 6truecm
\psfig{figure=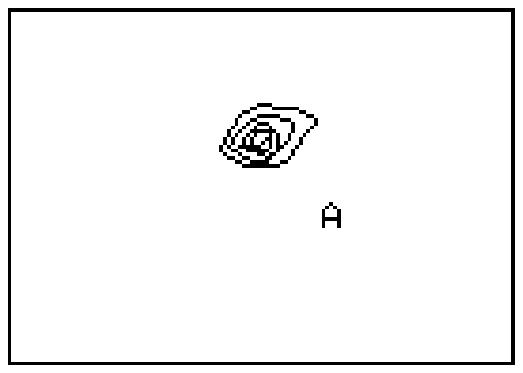,height=240mm,bbllx=70mm,bblly=60mm,bburx=140mm,bbury=250mm}
\vskip -7truecm
{\bf Figure 8:} Continuum (up), $H_\alpha$ map (middle) and zoomed isocontour 
map (down)  of HCG38. 
The lowest contour is at 1$\sigma$ ($7.22 \cdot 10^{-17}~erg~cm^2s^{-1}arcsec^{-2}$) above the background.
The interval is 1$\sigma$.
\end{figure*}
\newpage
\begin{figure*}[h]
\centerline{\psfig{figure=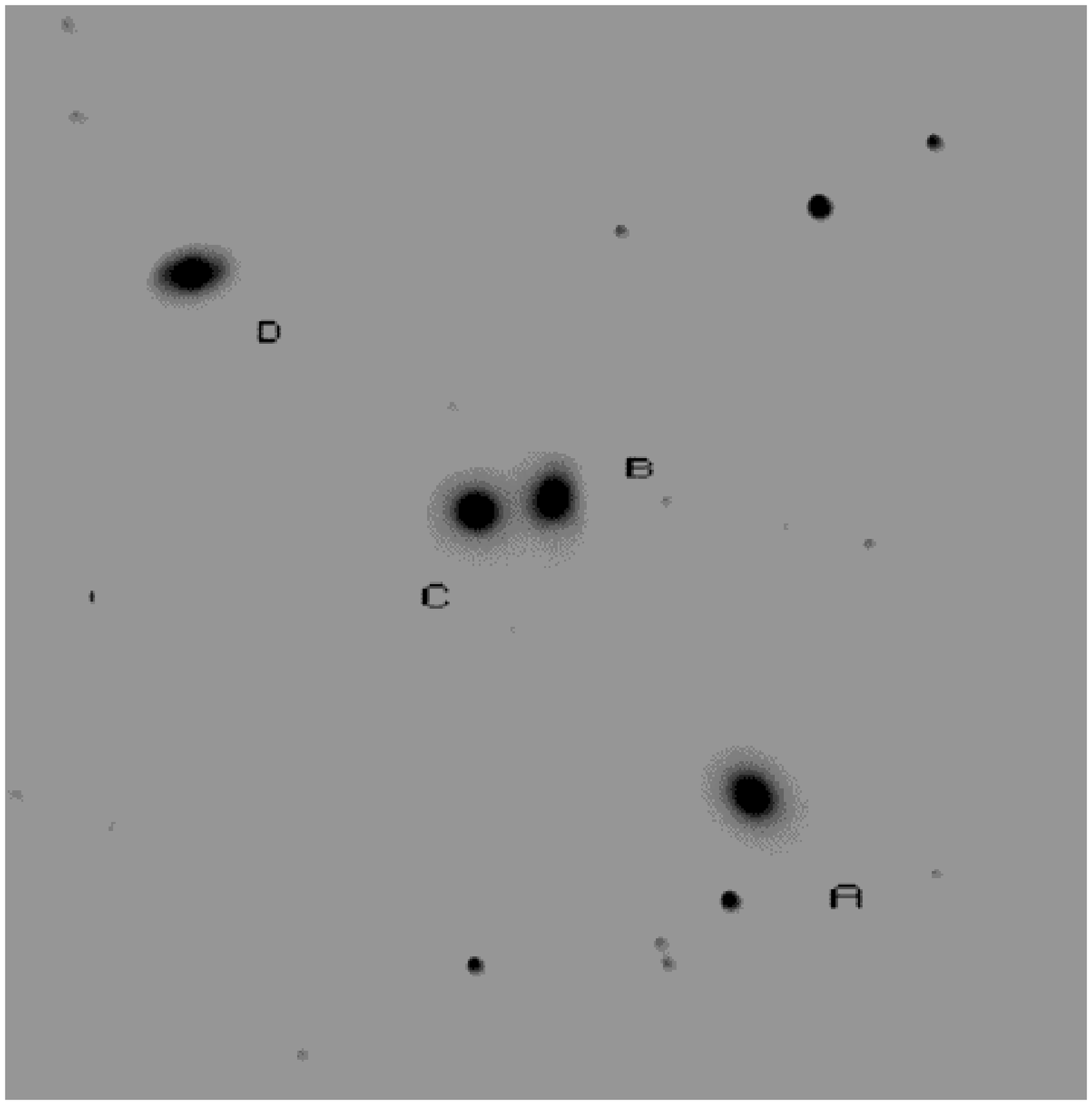,height=100mm,bbllx=70mm,bblly=60mm,bburx=140mm,bbury=250mm}}
\bigskip
\bigskip
\centerline{\psfig{figure=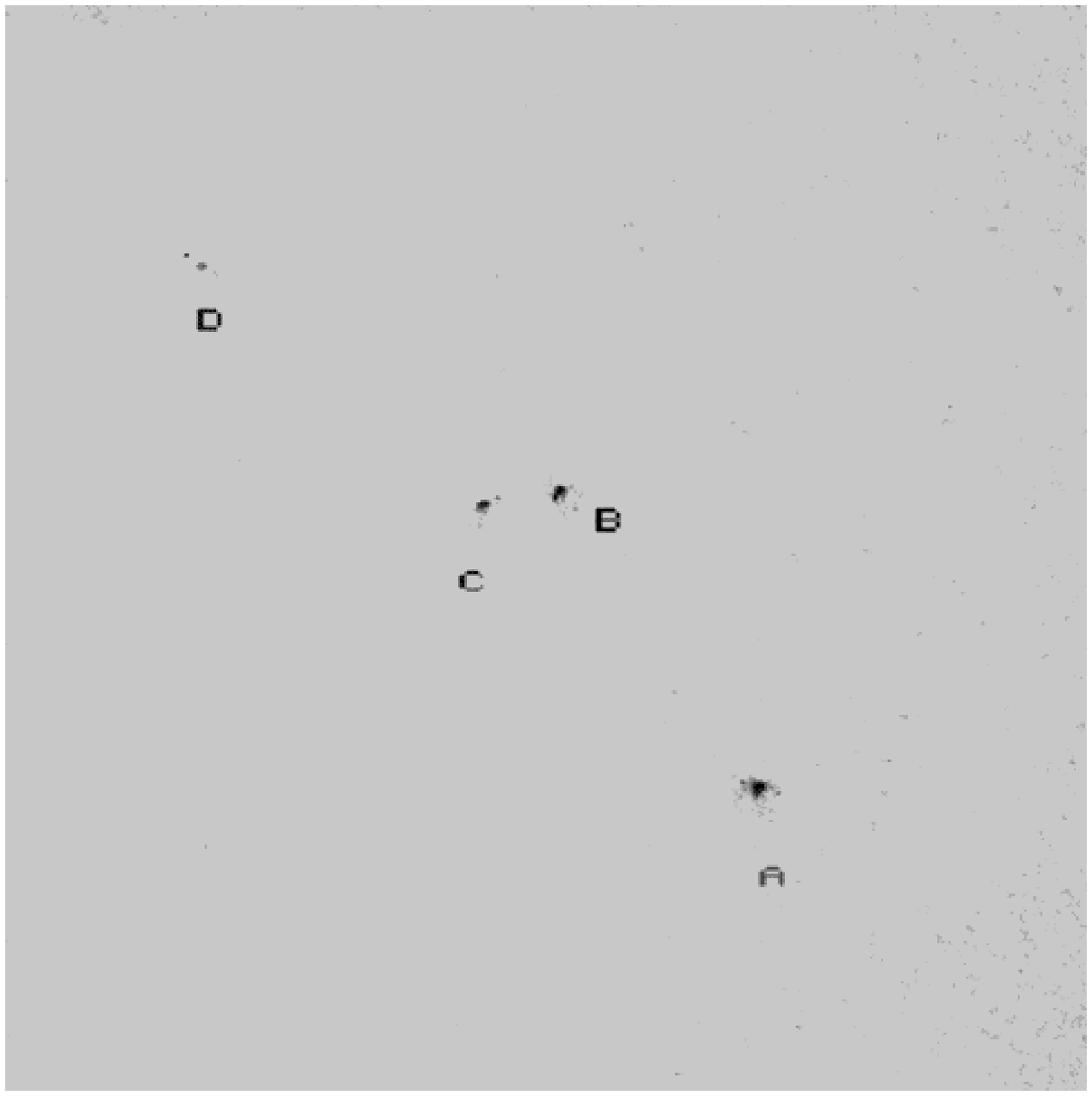,height=100mm,bbllx=70mm,bblly=60mm,bburx=140mm,bbury=250mm}}
\bigskip
\bigskip
{\bf Figure 9:} Continuum (up) and $H_\alpha$ map (down) of HCG46. 
\end{figure*}
\newpage
\begin{figure*}[h]
\centerline{\psfig{figure=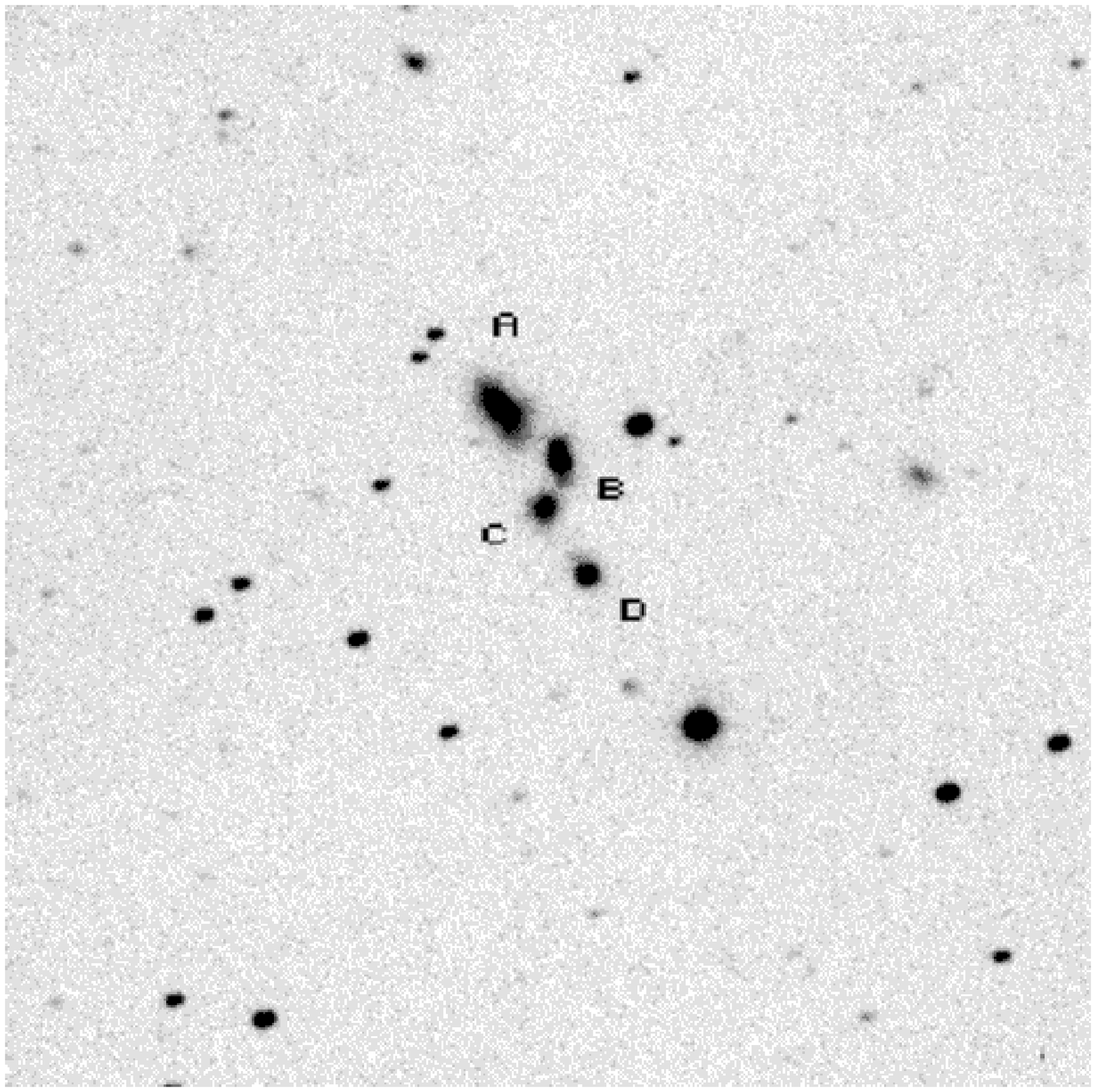,height=75mm,bbllx=70mm,bblly=60mm,bburx=140mm,bbury=250mm}}
\centerline{\psfig{figure=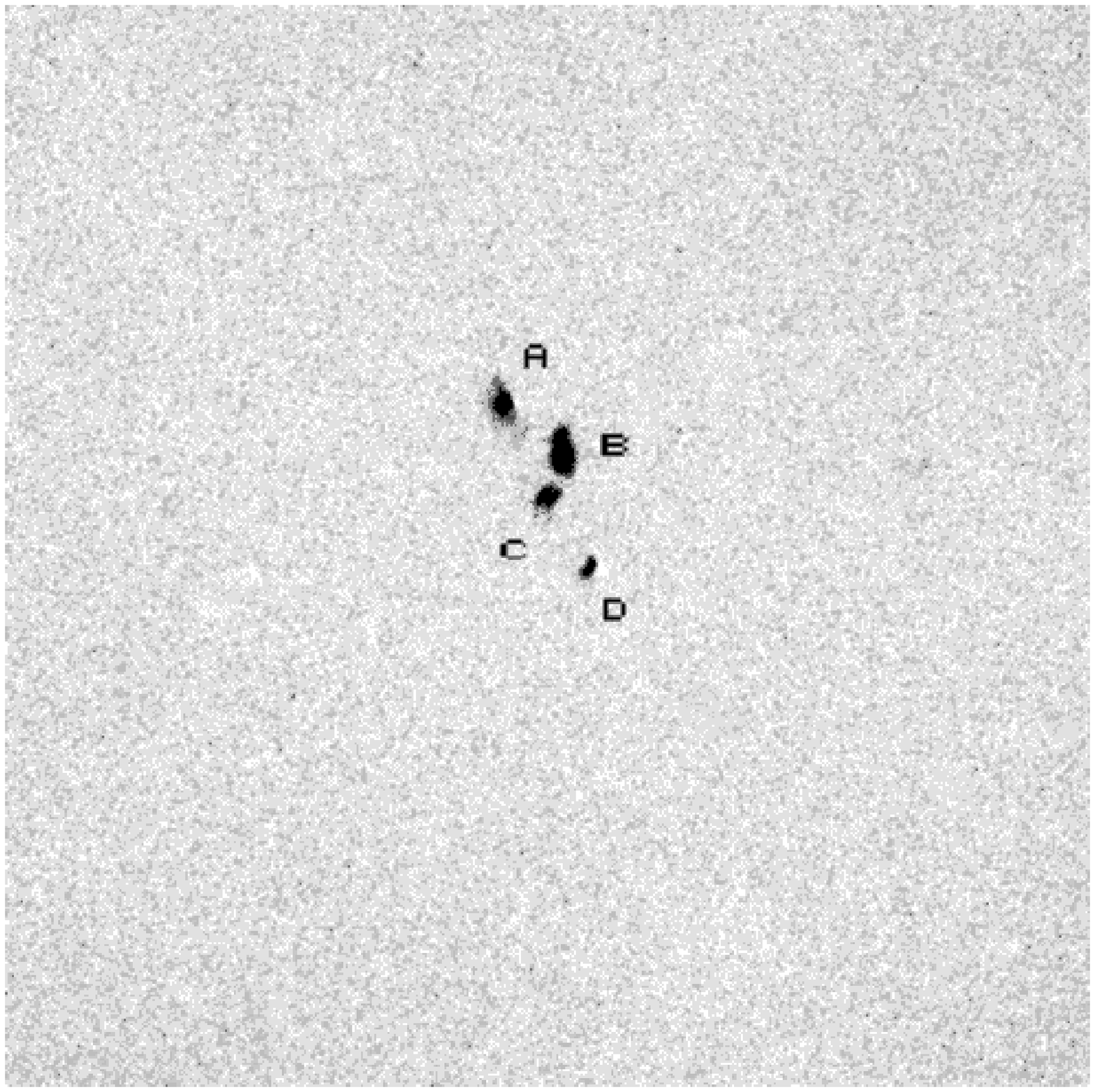,height=75mm,bbllx=70mm,bblly=60mm,bburx=140mm,bbury=250mm}}
\vskip -5.7truecm
\hskip -0.2truecm
\centerline{\psfig{figure=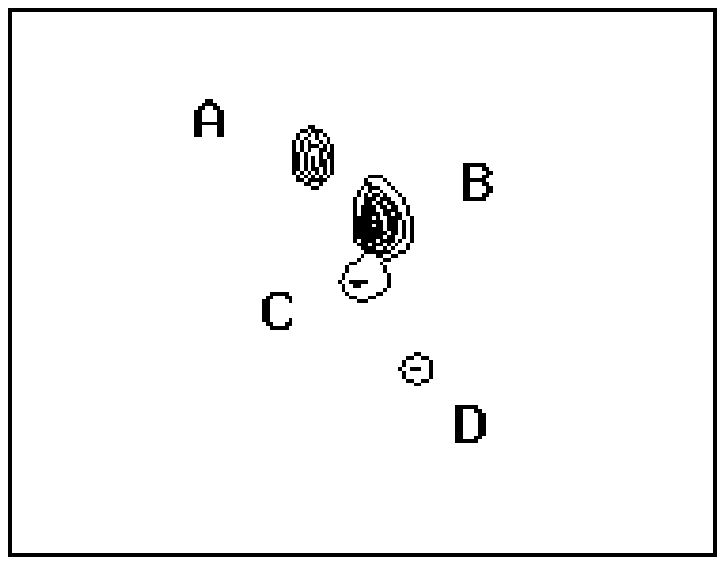,height=180mm,bbllx=70mm,bblly=60mm,bburx=140mm,bbury=250mm}}
\vskip -4.5truecm
{\bf Figure 10:} Continuum (up), $H_\alpha$ map (middle) and zoomed 
isocontour map (down) of HCG49. 
The lowest contour is at 1$\sigma$ ($9.44 \cdot 10^{-17}~erg~cm^2s^{-1}arcsec^{-2}$) above the background.
The interval among the contours is 2$\sigma$.
\end{figure*}
\newpage
\begin{figure*}[h]
\centerline{\psfig{figure=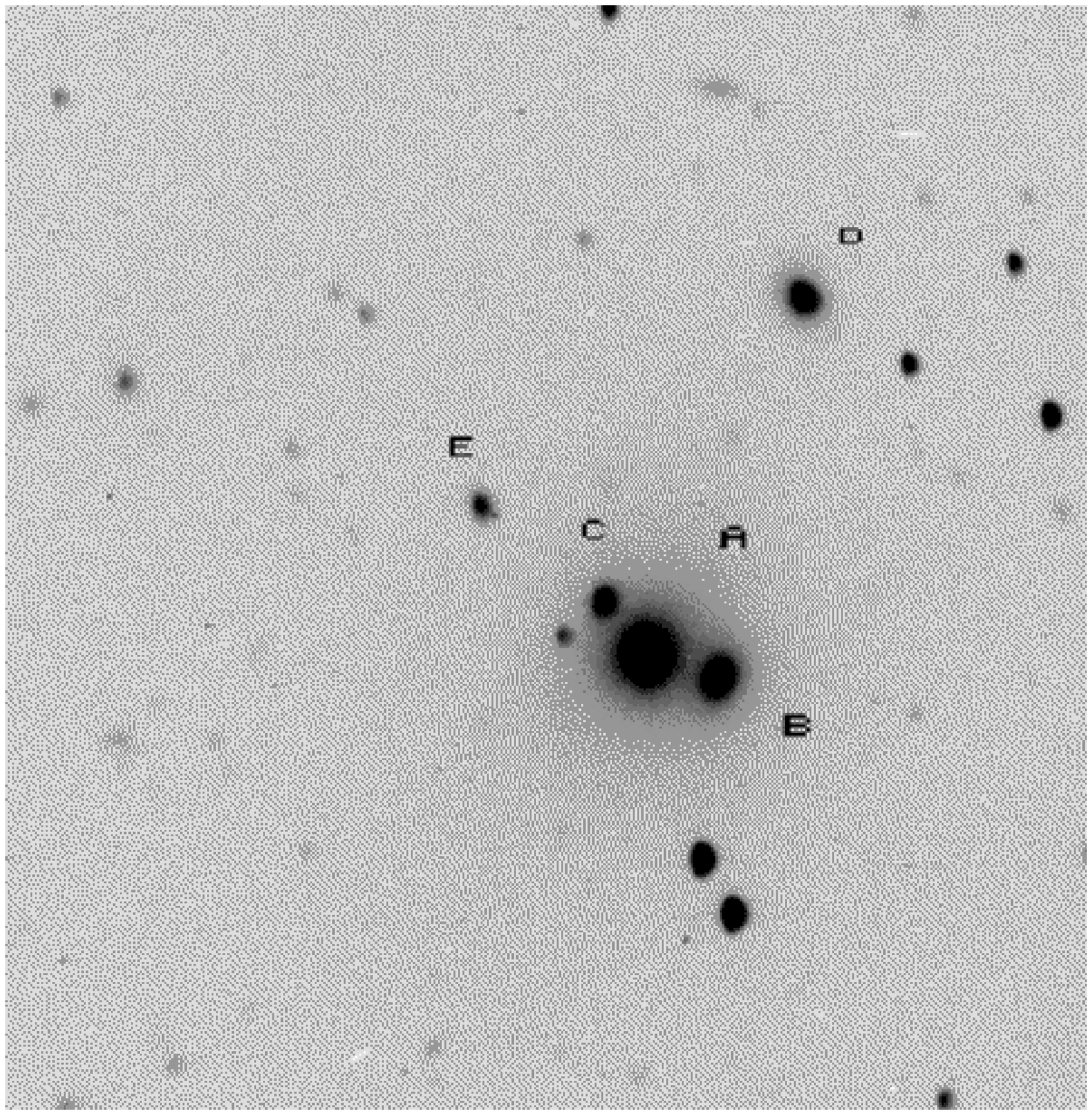,height=75mm,bbllx=70mm,bblly=60mm,bburx=140mm,bbury=250mm}}
\centerline{\psfig{figure=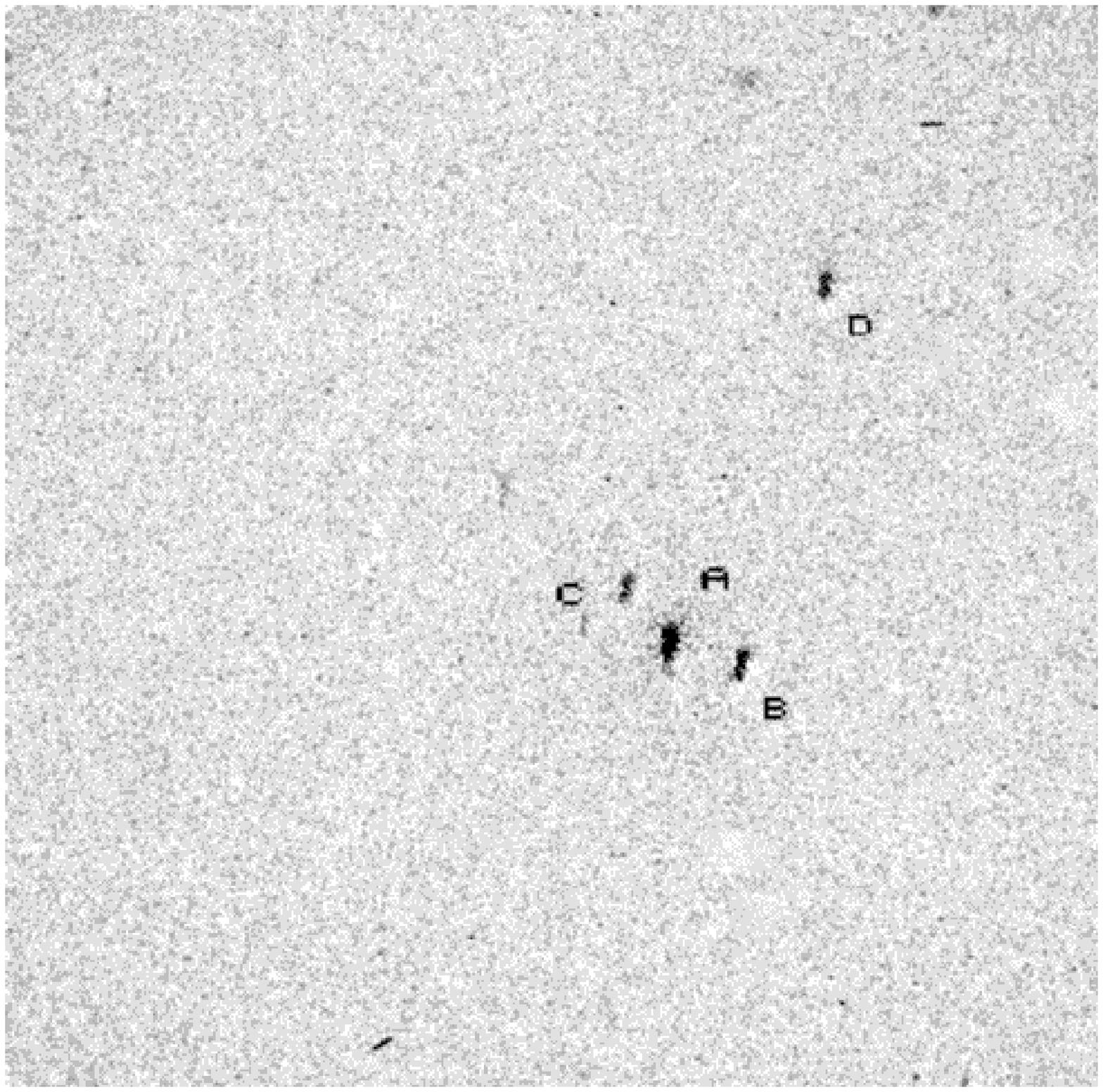,height=75mm,bbllx=70mm,bblly=60mm,bburx=140mm,bbury=250mm}}
\vskip -5.7truecm
\hskip -0.25truecm
\centerline{\psfig{figure=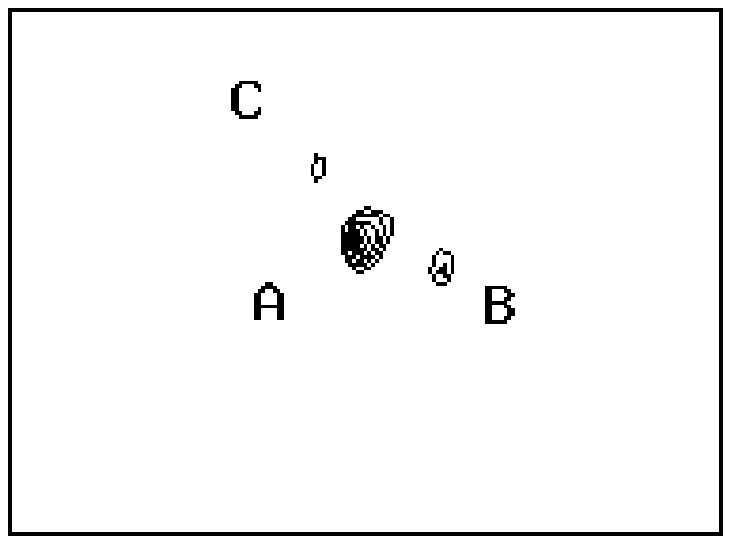,height=180mm,bbllx=70mm,bblly=60mm,bburx=140mm,bbury=250mm}}
\vskip -5truecm
{\bf Figure 11:} Continuum (up), $H_\alpha$ map (middle) and zoomed 
isocontour map (down) of HCG74. 
The isocontour plots are given for A, B and C  
galaxies.
The lowest contour is at 1$\sigma$ ($7.52 \cdot 10^{-17}~erg~cm^2s^{-1}arcsec^{-2}$) above the background.
The interval among the contours is 1$\sigma$.
\end{figure*}

%   \begin{figure*}
%     \vspace{4cm}
%\rule{0.4pt}{4cm}% line thickness, height of picture
%     \caption{Adiabatic exponent $\Gamma_1$}
%\hfill      \parbox[b]{5cm}{\caption[]{Adiabatic exponent $\Gamma_1$.
%               $\Gamma_1$ is plotted as a function of
%               $\lg$ internal energy $\mathrm{[erg\,g^{-1}]}$ and $\lg$
%               density $\mathrm{[g\,cm^{-3}]}$
%              }}%
%         \label{FigGam}%
%    \end{figure*}
%

%__________________________________________________ One column table
%   \begin{table}
%      \caption[]{Opacity sources}
%         \label{KapSou}
%      \[
%         \begin{array}{p{0.5\linewidth}l}
%            \hline
%            \noalign{\smallskip}
%            Source      &  T / {[\mathrm{K}]} \\
%            \noalign{\smallskip}
%            \hline
%            \noalign{\smallskip}
%            Yorke 1979, Yorke 1980a & \leq 1700^{\mathrm{a}}     \\
%           Yorke 1979, Yorke 1980a & \leq 1700             \\
%            Kr\"ugel 1971           & 1700 \leq T \leq 5000 \\
%            Cox \& Stewart 1969     & 5000 \leq             \\
%            \noalign{\smallskip}
%            \hline
%         \end{array}
%      \]
%\begin{list}{}{}
%\item[$^{\mathrm{a}}$] This is footnote a
%\end{list}
%   \end{table}
%
%
%___________________________________ Two column table (place early!)

\begin{acknowledgements}
We are grateful to E. Recillas and D. Maccagni which have kindly carried out one of the observing runs. We would like also to thank the S. Pedro Martir observatory staff for the helpful support given during  observations. PS would like to acknowledge and to thank B. Catinella and S. Molendi for the useful discussions and suggestions given during this work was in progress.                                                            
\end{acknowledgements}

\end{document}